\pdfoutput=1
\documentclass[letterpaper]{article}
\usepackage{geometry}
\usepackage[T1]{fontenc}
\usepackage{newtxtext}
\usepackage{helvet}
\usepackage{courier}
\usepackage{etoolbox}
\AtBeginEnvironment{table}{\setlength{\belowcaptionskip}{6pt}}
\makeatletter
\let\@dblfloat\@float
\let\end@dblfloat\end@float
\makeatother

\usepackage[T1]{fontenc}
\usepackage[utf8]{inputenc}
\usepackage{microtype}
\usepackage{amsmath,amssymb}
\usepackage{graphicx}
\usepackage{xcolor}
\usepackage{booktabs}
\usepackage{longtable}
\usepackage{float}
\usepackage{array}
\usepackage{listings}
\usepackage{upquote}
\usepackage{algorithm}
\usepackage[noend]{algpseudocode}
\usepackage{xspace}
\usepackage{tikz}
\usetikzlibrary{positioning,arrows.meta,calc,fit,backgrounds,decorations.pathreplacing}
\usepackage[hyphens]{url}
\usepackage[hidelinks]{hyperref}
\hypersetup{pdftitle={count-catcher: Lean Theorems from Model Counting Certificates},
  pdfauthor={Johannes K. Fichte, Stefan Szeider},
  pdfkeywords={model counting, proof certificates, Lean 4, reflection}}
\usepackage{cleveref}
\crefname{section}{Section}{Sections}\Crefname{section}{Section}{Sections}
\crefname{table}{Table}{Tables}\crefname{figure}{Figure}{Figures}
\crefname{lstlisting}{Listing}{Listings}\Crefname{lstlisting}{Listing}{Listings}
\crefname{appendix}{Appendix}{Appendices}\crefname{subappendix}{Appendix}{Appendices}

\usepackage[scaled=0.85]{sourcecodepro}
\lstdefinelanguage{lean}{
  morekeywords={theorem,def,lemma,example,where,let,in,if,then,else,match,with,
    do,return,have,show,by,sorry,import,open,namespace,end,section,variable,
    instance,class,structure,inductive,noncomputable,deriving,extends,abbrev,
    axiom,private,protected,mutual,attribute,macro,syntax,notation,Type,Prop,
    Sort,true,false,native_decide,decide},
  sensitive=true,
  morecomment=[l]{--},
  morecomment=[s]{/-}{-/},
  morestring=[b]",
}
\newcommand{\lean}{\lstinline[breaklines=true,breakatwhitespace=true]}

\crefname{listing}{Listing}{Listings}\Crefname{listing}{Listing}{Listings}
\crefname{algorithm}{Algorithm}{Algorithms}\Crefname{algorithm}{Algorithm}{Algorithms}
\AtBeginDocument{}
\newcommand{\tool}{count-catcher\xspace}

\newcommand{\funTotal}{180}
\newcommand{\funPerTrack}{100}
\newcommand{\funDuplicates}{20}

\newcommand{\funLimitDFour}{4{,}000}
\newcommand{\funLimitGen}{4{,}000}
\newcommand{\funLimitLean}{3{,}600}
\newcommand{\funLimitMem}{64}
\newcommand{\funDFourEntered}{180}
\newcommand{\funDFourPassed}{123}
\newcommand{\funDFourTimeout}{57}
\newcommand{\funDFourOom}{0}
\newcommand{\funDFourCrash}{0}
\newcommand{\funGenEntered}{123}
\newcommand{\funGenPassed}{97}
\newcommand{\funGenTimeout}{22}
\newcommand{\funGenOom}{4}
\newcommand{\funGenCrash}{0}
\newcommand{\funTransEntered}{97}
\newcommand{\funTransPassed}{97}
\newcommand{\funTransTimeout}{0}
\newcommand{\funTransOom}{0}
\newcommand{\funTransCrash}{0}
\newcommand{\funLeanEntered}{97}
\newcommand{\funLeanPassed}{96}
\newcommand{\funLeanRejected}{0}
\newcommand{\funLeanTimeout}{1}
\newcommand{\funLeanOom}{0}
\newcommand{\funLeanCrash}{0}
\newcommand{\funProved}{96}

\newcommand{\funDFourMed}{3.5}

\newcommand{\funGenMed}{30.3}

\newcommand{\funLeanMed}{15.7}

\newcommand{\bigCertGB}{33.9}

\newcommand{\bigCountDigits}{1{,}681}

\newcommand{\funLeanTimeoutCertGB}{126.9}

\newcommand{\cmpN}{91}
\newcommand{\cmpNMcTwentyTwentyTwo}{82}
\newcommand{\cmpNFamilies}{9}

\newcommand{\cmpRuleGenLimit}{600}
\newcommand{\cmpOursProved}{91}
\newcommand{\cmpVcCorrect}{91}

\newcommand{\cmpCcCorrect}{84}
\newcommand{\cmpCcWrong}{7}

\newcommand{\cmpOurCheckMed}{4.5}

\newcommand{\cmpFwdSharePct}{77}
\newcommand{\cmpFwdShareN}{89}

\newcommand{\cmpVcMed}{22.6}

\newcommand{\cmpCcMed}{4.1}

\newcommand{\cmpOurMemMed}{3.4}
\newcommand{\cmpOurMemMin}{3.2}

\newcommand{\cmpVcMemMed}{0.99}

\newcommand{\cmpCcMemMed}{0.095}

\newcommand{\cmpBoundCheckMed}{0.23}

\newcommand{\cmpBoundCheckN}{91}
\newcommand{\cmpBoundCheckLower}{91}
\newcommand{\cmpBoundThmMed}{1.2}

\newcommand{\cmpBoundThmN}{91}
\newcommand{\cmpBoundThmLower}{38}
\newcommand{\cmpBoundPipeMed}{0.92}

\newcommand{\cmpBoundPipeN}{91}
\newcommand{\cmpBoundPipeLower}{55}
\newcommand{\cmpBoundMemMed}{3.6}

\newcommand{\cmpBoundMemN}{91}
\newcommand{\cmpBoundMemLower}{14}
\newcommand{\cmpThmOursMed}{22.3}
\newcommand{\cmpPipeOursMed}{54.5}
\newcommand{\cmpPipeVcMed}{88.2}
\newcommand{\cmpPipeCcMed}{65.5}

\newcommand{\cmpBandOneLabel}{below 10\,MB}
\newcommand{\cmpBandOneN}{18}
\newcommand{\cmpBandOneLeanRatio}{86}
\newcommand{\cmpBandOneMemRatio}{37}
\newcommand{\cmpBandOneCheckRatio}{0.22}
\newcommand{\cmpBandTwoLabel}{10 to 100\,MB}
\newcommand{\cmpBandTwoN}{15}
\newcommand{\cmpBandTwoLeanRatio}{3.9}
\newcommand{\cmpBandTwoMemRatio}{12}
\newcommand{\cmpBandTwoCheckRatio}{0.28}
\newcommand{\cmpBandThreeLabel}{100\,MB to 1\,GB}
\newcommand{\cmpBandThreeN}{43}
\newcommand{\cmpBandThreeLeanRatio}{0.93}
\newcommand{\cmpBandThreeMemRatio}{2.5}
\newcommand{\cmpBandThreeCheckRatio}{0.20}
\newcommand{\cmpBandFourLabel}{above 1\,GB}
\newcommand{\cmpBandFourN}{15}
\newcommand{\cmpBandFourLeanRatio}{0.34}
\newcommand{\cmpBandFourMemRatio}{0.53}
\newcommand{\cmpBandFourCheckRatio}{0.22}

\newcommand{\cmpBigInstance}{track2\_011}
\newcommand{\cmpBigFullGB}{6.7}

\newcommand{\cmpBigOurLean}{184.7}
\newcommand{\cmpBigOurMem}{6.8}
\newcommand{\cmpBigVc}{519.9}
\newcommand{\cmpBigVcMem}{24.2}

\newcommand{\ablRevBytesMed}{2.0}
\newcommand{\ablRevBytesMin}{1.1}
\newcommand{\ablRevBytesMax}{17}
\newcommand{\ablRevBytesN}{91}
\newcommand{\ablRevGenMed}{1.5}
\newcommand{\ablRevGenMin}{0.97}
\newcommand{\ablRevGenMax}{42}
\newcommand{\ablRevGenN}{75}

\newcommand{\ablPartcTdMed}{4.6}
\newcommand{\ablPartcTdMin}{1.0}
\newcommand{\ablPartcTdMax}{17}
\newcommand{\ablPartcTdN}{17}

\newcommand{\ablPartcPcMed}{90}
\newcommand{\ablPartcPcMin}{41}
\newcommand{\ablPartcPcMax}{220}
\newcommand{\ablPartcPcN}{15}
\newcommand{\ablPartcPcCensored}{2}

\newcommand{\ablTrimBinPartBMed}{1.4}
\newcommand{\ablTrimBinPartBMin}{0.71}
\newcommand{\ablTrimBinPartBMax}{1.9}
\newcommand{\ablTrimBinPartBN}{7}

\newcommand{\ablKeepLeanMed}{1.2}
\newcommand{\ablKeepLeanMin}{1.1}
\newcommand{\ablKeepLeanMax}{1.9}
\newcommand{\ablKeepLeanN}{7}
\newcommand{\ablKeepMemMed}{1.7}
\newcommand{\ablKeepMemMin}{1.0}
\newcommand{\ablKeepMemMax}{2.0}
\newcommand{\ablKeepMemN}{7}
\newcommand{\ablInlinePartBMed}{1.7}
\newcommand{\ablInlinePartBMin}{1.1}
\newcommand{\ablInlinePartBMax}{2.9}
\newcommand{\ablInlinePartBN}{7}

\newcommand{\ablInlineTransMed}{0.73}
\newcommand{\ablInlineTransMin}{0.70}
\newcommand{\ablInlineTransMax}{0.88}
\newcommand{\ablInlineTransN}{7}

\newcommand{\ablAllPartBMed}{3.2}
\newcommand{\ablAllPartBMin}{1.6}
\newcommand{\ablAllPartBMax}{4.2}
\newcommand{\ablAllPartBN}{7}

\newcommand{\ablKeepShareMed}{33}
\newcommand{\ablKeepShareMin}{18}
\newcommand{\ablKeepShareMax}{35}
\newcommand{\ablKeepShareN}{7}

\newcommand{\ablEmbedSameMed}{0.90}
\newcommand{\ablEmbedSameMin}{0.82}
\newcommand{\ablEmbedSameMax}{0.93}
\newcommand{\ablEmbedSameN}{11}
\newcommand{\ablEmbedDefMed}{1.9}
\newcommand{\ablEmbedDefMin}{1.4}
\newcommand{\ablEmbedDefMax}{2.1}
\newcommand{\ablEmbedDefN}{11}

\newcommand{\kerFourAdds}{99}
\newcommand{\kerFourNativeMs}{16}
\newcommand{\kerFourKernelS}{24}

\newcommand{\kerLimitGB}{64}

\newcommand{\unshN}{91}
\newcommand{\unshCheckMed}{0.39}

\newcommand{\unshCheckLower}{91}

\newcommand{\rgFailN}{26}
\newcommand{\rgGenOk}{10}
\newcommand{\rgThm}{8}
\newcommand{\rgThmLeanTenK}{10}

\newcommand{\rgTotal}{104}

\newcommand{\scTriTenCount}{12{,}172}

\newcommand{\scTriTenComposed}{42}
\newcommand{\scTriTenLean}{74.8}

\newcommand{\scLatinSevenCountSci}{\ensuremath{6.15\cdot 10^{13}}}
\newcommand{\scLatinSevenComposed}{4}
\newcommand{\scLatinSevenLean}{64.0}

\newcommand{\scDedekindSevenCountSci}{\ensuremath{2.41\cdot 10^{12}}}
\newcommand{\scDedekindSevenComposed}{1}
\newcommand{\scDedekindSevenLean}{88.8}

\newcommand{\scKrewerasSixCountSci}{\ensuremath{2.06\cdot 10^{17}}}
\newcommand{\scKrewerasSixComposed}{1}
\newcommand{\scKrewerasSixLean}{170.5}

\newcommand{\scSpernerSevenCountSci}{\ensuremath{6.96\cdot 10^{11}}}
\newcommand{\scSpernerSevenComposed}{1}
\newcommand{\scSpernerSevenLean}{424.6}

\newcommand{\scTotDomNineCountSci}{\ensuremath{7.27\cdot 10^{22}}}
\newcommand{\scTotDomNineComposed}{1}
\newcommand{\scTotDomNineLean}{145.0}

\newcommand{\scTotDomTenCountSci}{\ensuremath{1.68\cdot 10^{28}}}
\newcommand{\scTotDomTenComposed}{1}
\newcommand{\scTotDomTenLean}{14.8}

\newcommand{\scRamseyThirteenCount}{1}

\newcommand{\scRamseyThirteenComposed}{113}
\newcommand{\scRamseyThirteenLean}{78.9}

\newcommand{\scLatinEightCount}{108{,}776{,}032{,}459{,}082{,}956{,}800}

\newcommand{\scLatinEightCountSci}{\ensuremath{1.09\cdot 10^{20}}}

\newcommand{\scLatinEightLeaves}{737}
\newcommand{\scLatinEightComposed}{737}
\newcommand{\scLatinEightLean}{325{,}252}
\newcommand{\scLatinEightLeanHours}{90}

\newcommand{\scLatinEightNativeAxioms}{1{,}474}
\newcommand{\scLatinEightCertTB}{7.9}

\newcommand{\scLatinEightCertMaxGB}{32.7}

\newcommand{\famN}{11}

\newcommand{\famLatinLines}{983}
\newcommand{\famKrewerasLines}{499}

\newcommand{\miscountFormula}{random cubic, 100 vertices}
\newcommand{\miscountOurs}{8{,}281{,}581{,}627{,}129{,}350{,}099}
\newcommand{\miscountPrinted}{8{,}281{,}581{,}627{,}129{,}036{,}755}

\begin{document}

\title{count-catcher: Lean Theorems from\\ Model Counting Certificates}

\author{Johannes K. Fichte\\[4pt]
  \small Centre de Recherche en Informatique de Lens (CRIL), UMR 8188\\[-3pt]
  \small Universit\'e d'Artois and CNRS, Lens, France
  \and
  Stefan Szeider\\[4pt]
  \small Algorithms and Complexity Group\\[-3pt]
  \small TU Wien, Vienna, Austria\\[-3pt]
  \small \href{https://www.ac.tuwien.ac.at/people/szeider/}{www.ac.tuwien.ac.at/people/szeider/}
}
\date{}

\maketitle
\thispagestyle{empty}

\begin{abstract}
We present \tool, a tool that turns a CPOG certificate of a count by the exact model counter d4 into a Lean theorem that a CNF formula defined in Lean has exactly~$N$ models over~$V$ variables.
With an encoding lemma, this theorem becomes a theorem about the number of combinatorial objects.
A CPOG certificate consists of a Boolean circuit equivalent to the formula and a proof of equivalence.
\tool checks in Lean that the formula implies the circuit (forward check) and the circuit implies the formula (reverse check).
For the forward check, lrat-catcher streams a refutation through the LRAT checker of Lean's standard library.
The reverse check and the validity check of the circuit are computations proved correct in Lean.
Lean runs all three checks as compiled code, so the theorem depends on Lean's kernel, compiler and runtime, as does CPOG's verified checker.
We combine the theorem with results from Mathlib, for instance with Burnside's lemma to count graphs up to isomorphism.
We prove the number of Latin squares of order~8 as a weighted sum of hundreds of certified counts from about eight terabytes of certificates.
We turn about half of the 2022 Model Counting Competition formulas into theorems, with the three checks taking about a quarter of the time of CPOG's verified checker.
We found that CPOG's unverified checker reported wrong counts, and removed its verified checker's dependence on two unproved lemmas.
End to end, a theorem costs about the same as a checked count.

\end{abstract}

\section{Introduction}\label{sec:intro}

There are exactly \scLatinEightCount{} Latin squares of order~8.
Numbers like this are established by computation.
A common approach is to encode the objects as models of a propositional formula and run an exact model counter.
For the Latin squares, the formula has one variable per cell and symbol, and its clauses ensure that every row and column forms a permutation.
In Lean, we state the number using the following definition of a Latin square:
\begin{lstlisting}
def IsLatin {n : ℕ} (L : Fin n → Fin n → Fin n) : Prop :=
  (∀ i, Function.Bijective (L i)) ∧
    ∀ j, Function.Bijective fun i => L i j
theorem latin8_all :
    Nat.card {L : Fin 8 → Fin 8 → Fin 8 // IsLatin L}
      = 108776032459082956800
\end{lstlisting}
 The first thing we need to prove is an \emph{encoding lemma} that ensures that the Latin squares are in bijection with the models of the formula.
The user provides this lemma.
The second thing is the \emph{count theorem}, that a CNF formula~$F$ given as a Lean term has exactly~$N$ models over a universe of~$V$ variables that contains the variables of~$F$.
The following is the soundness theorem of our checker:
\begin{lstlisting}
theorem checkAll_sound (h : checkAll V F pogStr lratStr N = true) :
    Fintype.card {a : Fin V → Bool // F.eval (ext a) = true} = N
\end{lstlisting}
 The listing shows that the hypothesis \lean{h} says that the checker accepts a certificate, given as the strings \lean{pogStr} and \lean{lratStr}, and the part after the colon is the count theorem.
So once the checker has accepted, the count theorem follows from the soundness theorem.
Here, \lean{ext a} extends the assignment~$a$ of the universe by false on all other variables.
From the encoding lemma and the count theorem, we obtain an \emph{end-to-end} theorem in which the number is verified inside Lean from the definition of the objects to the final number.
How to establish such a count theorem at the scale of terabytes of certificates is the subject of this paper.

There is a long history of claims of counts that turned out to be wrong, which emphasizes the importance of machine-checkable proofs.
For instance, Sade corrected Norton's 1939 count of the reduced Latin squares of order~7 in 1948~\cite{Norton39,Sade48,Sade51,McKayW05}, Berman and K\"ohler confirmed Church's 1965 value of Dedekind's number $M(7)$, the number of monotone Boolean functions of seven variables, against Lunnon's of 1971~\cite{Church65,Lunnon71,BermanK76}, and McKay corrected the 1996 count of the closed knight's tours of the chessboard a year later~\cite{LobbingW96,McKay97}.
An independent program that checks a certificate excludes such errors only if it is itself correct.
In our evaluation (\cref{sec:eval}), cpog-check, the unverified checker of the certificate format CPOG (certified partitioned-operation graphs), reported a wrong count on \cmpCcWrong{} of \cmpN{} certificates.

Model counting, determining the number of models of a propositional formula, is a fundamental \#P-complete problem~\cite{Valiant79}.
Exact counters~\cite{GomesSS21} output only the integer of the count, so an error in their search cannot be detected from the output.
In certified model counting, the counter or a separate generator also outputs a certificate, which can be checked by an independent checker.
CPOG~\cite{BryantNAH23,BryantNAH25} consists of a certificate format for the circuits that counters such as d4~\cite{LagniezM17} generate, a generator for it, and a checker whose correctness is proved in Lean~\cite{MouraU21}.
Such a correctness theorem for the checker is about the checker's final state.
It certifies that if a run ends in the final state, the output count is the count of the parsed DIMACS file (the standard text format of CNF formulas).
However, there is no \emph{theorem} that connects this count to a Lean term or to the combinatorial objects that are counted, like the Latin squares in our example.
A different approach is to \emph{enumerate} the objects and use proof logging in VeriPB and CakePB~\cite{McCreeshNOT26} to obtain theorems in the proof assistant HOL4 about objects.
However, enumeration is not a feasible approach if the number of objects is in the order we are considering in this paper.

We present \tool, which takes a CNF formula given as a Lean term and returns its model count as a Lean theorem.
The user provides the encoding lemma, and our tool exports the formula and runs the exact counter d4 and the CPOG generator.
With an untrusted translator, it translates the CPOG certificate into a node table that contains the circuit and an LRAT refutation, which it may shorten.
It then writes one line, \lean{count_models}, into a new Lean theorem file.
This line names the theorem, the encoding lemma to apply, the certificate files and the count.
When Lean processes this file, it runs a Lean-verified checker on the certificate.
Lean then proves the count theorem, an instance of \lean{checkAll_sound}, and from it and the encoding lemma we obtain the end-to-end theorem, for example \lean{latin8_all}.

Count theorems can then be used in larger proofs, for instance, to count only up to symmetry.
This is made possible by means of Burnside's lemma~\cite[Chapter~2]{HararyP73}, which counts objects up to symmetry as the average number of objects that a symmetry fixes.
Using this, we can show, for instance, that the number of triangle-free graphs on ten vertices up to isomorphism is a weighted sum of 42 counts of labeled graphs, one for each cycle type (the multiset of cycle lengths) of a permutation of the ten vertices.
Here, the count for a cycle type is the number of triangle-free graphs that a permutation of this type fixes.
Each of the 42 numbers is a separate model count of a formula.
Each count is shown as a theorem with \tool.
The remainder of the proof uses Burnside's lemma from Mathlib~\cite{Mathlib20}, Lean's mathematical library, and a decision procedure that checks that the list of cycle types is complete and that the weighted sum gives the claimed number (\cref{sec:burnside}).
We count the Latin squares of order~8 in the same way, as a weighted sum over the cycle types of the second row, which in a reduced square (first row and first column are the identity) is a permutation of the symbols.
Counting up to symmetry also allows us to obtain uniqueness results.
Up to isomorphism, exactly one graph on 13 vertices has neither a triangle nor five mutually nonadjacent vertices (a $(3,5)$-Ramsey graph)~\cite{Kery64,Radziszowski}.
From a count of one isomorphism class, composed of certified counts, we prove in Lean that every such graph is isomorphic to the circulant graph with jumps 1 and~5.

The difficulty is to bring the equality $\#_V(F)=N$ into the logic at the scale of the certificates.
Here, $\#_V(F)$ is the number of models of~$F$ over the universe.
To use the count as a lemma about the objects, the theorem must be about the formula~$F$ with its clauses and universe.
Unlike unsatisfiability, which is preserved when clauses or variables are added, the count can change with one clause more or one variable more.
The certificates of the equality in cases of interest are large.
We use one certificate for each cycle type of the second row of the Latin square.
These are four for order~7 and seven for order~8.
There we split the squares of each cycle type further by three cells of the third row into \scLatinEightLeaves{} certificates.
For the Latin squares of order~7, the certificates take under a gigabyte in total.
For order~8, they take \scLatinEightCertTB~TB, with up to \scLatinEightCertMaxGB~GB each (\cref{app:showcases}).
Certificates of this size are too large for a single proof term, and the kernel cannot process them.

The equality is shown by combining the encoding lemma, one refutation and two computations proved correct in Lean, forming the chain
\begin{equation}\label{eq:chain}
  |X|\;\overset{\text{encoding}}{=}\;\#_V(F)\;\overset{\text{forward, reverse}}{=}\;\#_V(P)\;\overset{\text{validity}}{=}\;c(r)\cdot 2^{\,V-|S(r)|}\;=\;N.
\end{equation}
Here, $X$ denotes the set of objects, for example the Latin squares of order~8, $F$ the formula that encodes them, and $P$ the circuit the certificate is about, with root~$r$, its output node.
The circuit $P$ is a POG (partitioned-operation graph), a directed acyclic graph whose leaves are literals and whose inner nodes are products (AND) and sums (OR) (\cref{sec:background}).
Here, $\#_V(P)$ denotes the number of assignments of the universe that make the root true, $S(r)$ the set of variables below it, and $c(r)$ the number of assignments of~$S(r)$ that make the root true.
The first equality is supported by the encoding lemma provided by the user.
The second is the \emph{forward implication}, that every model of~$F$ makes~$r$ true, together with the \emph{reverse implication}, that every assignment that makes~$r$ true is a model of~$F$.
The third is the validity of the POG: the arguments of every product have pairwise disjoint variable sets and the two arguments of every sum are not true together. Under these two conditions we compute its count bottom-up.
The fourth is arithmetic.
Following CPOG, we call the three checks the \emph{forward}, the \emph{reverse} and the \emph{validity check}.

Checking the reverse implication by evaluating the POG is based on the way it is proposed in SCPOG~\cite{BryantTH25}, which extends CPOG to projected counting (models that agree on a given set of variables count once).
The forward check requires the verification of one refutation, which proves the forward implication.
We shorten it with the translator (\cref{sec:partb}) and stream it into Lean by means of lrat-catcher~\cite{Szeider26a}.
The validity check and the reverse check are both proved correct in Lean.
We compose all in one theorem over the same term, POG and universe.
All checks are run by \emph{native evaluation}, i.e., Lean compiles the checker, executes it, and adds an axiom that the compiled checker returned true.
Our trusted base therefore contains, besides the Lean kernel, also the compiler and the runtime.
CPOG's verified checker uses the same trusted base.

Our contributions are as follows.
\begin{enumerate}
\item \emph{Count theorem}: we translate a CPOG certificate into a count theorem over a Lean term and its universe.
In this theorem, Lean builds everything that is mentioned; the certificate only supplies data that is checked.
We test our checker against enumeration on more than twelve million generated and mutated certificates (\cref{sec:trust}).
\item \emph{Bit-parallel reverse check}: we decide the reverse implication with a check, proved correct in Lean, that evaluates 63 clauses at once in the bits of a machine word.
It is \ablPartcTdMed{} times faster in the median than our Lean implementation of SCPOG's top-down evaluation (\cref{sec:partc}).
The translator shortens the refutation in three untrusted steps before the forward check (\cref{sec:partb}).
\item \emph{Counts as lemmas}: based on user-provided encoding lemmas, our tool establishes the count, and lemma and count together give an end-to-end theorem.
We demonstrate this with compositions by symmetry, such as the Latin squares of order~8, and with single counts, such as Dedekind's number $M(7)$.
Four of the single counts are values of sequences of the On-Line Encyclopedia of Integer Sequences (OEIS)~\cite{OEIS} that, to our knowledge, have not been listed before (\cref{sec:lemmas,sec:eval}).
\item \emph{Evaluation}: we compare our tool with CPOG's checkers on the 2022 Model Counting Competition~\cite{FichteH23} and turn \funProved{} of \funTotal{} formulas into theorems, with certificates of up to \bigCertGB~GB.
The three checks take about a quarter of the time of CPOG's verified checker (\cref{sec:eval}).
We found that cpog-check reports wrong counts, which a fix of one line of its source removes (\cref{app:miscount}), and that CPOG's verified checker depends on two unproved lemmas, a dependency we removed (\cref{sec:trust}).
\end{enumerate}

\paragraph{Code and experiments.}
\tool with its Lean proofs, the theorem files, the generators, our two patches of cpog-gen and the scripts behind every table and figure are at \url{https://github.com/leansolving/count-catcher}; the Zenodo record \href{https://doi.org/10.5281/zenodo.23304375}{10.5281/zenodo.23304375} archives this tree at the release together with the prebuilt tools for x86-64, the logs and result files of the cluster runs, and the list of the competition formulas with their SHA-256 hashes.

\section{Related Work}\label{sec:related}

Bryant et al.~\cite{BryantNAH23,BryantNAH25} introduced CPOG, which includes the certificate format that we use, the generator cpog-gen and a Lean-verified checker, cpog-lean-check.
In \cref{tab:neighbours}, we compare its checkers, related checkers and our tool in terms of what a run ends in, what is verified and what is trusted.
Bryant, Tan and Heule~\cite{BryantTH25} introduced SCPOG, which extends CPOG's format to projected counting.
Each SCPOG certificate has no deletions (the lines that prove the reverse implication, \cref{sec:background}), since the reverse implication can be decided from the circuit alone~\cite{Capelli19,CapelliLM21,BeyersdorffFHHK26}.
Its checker is verified in HOL4 and compiled with CakeML.
It guarantees that the formula and the circuit agree after projection, but it does not guarantee any count obtained from this circuit.
The proof systems MICE~\cite{FichteHR22} and CD4~\cite{CapelliLM21} come with checkers that are not verified.
Beyersdorff, Hoffmann and Kasche~\cite{BeyersdorffHK25} study CD4's certificates as annotated decision-DNNFs and show that, with slight adaptations, they are equivalent to MICE.
Beyersdorff et~al.~\cite{BeyersdorffFHHK26} compare MICE with the knowledge compilation proof system kcps~\cite{Capelli19} in terms of proof complexity and show that CPOG is strictly stronger than MICE and kcps: it simulates both and is exponentially more succinct on some formulas. Chede, Chew and Shukla~\cite{ChedeCS24} introduced the proof system CLIP, which simulates kcps, CPOG and MICE.
Tan et al.~\cite{TanYSMM24} introduced ApproxMCCert, which certifies approximate model counts.
Proof logging with VeriPB and CakePB~\cite{McCreeshNOT26} takes an enumeration approach.
The solver logs every solution it finds, the verified checker checks the log, and the result is a HOL4 theorem about the objects.
McCreesh et al. certify enumeration and count-preserving transformations, but leave counting without enumeration open.
Since the log contains all solutions, the enumeration approach is not feasible for large counts as in our Latin squares count of order~8 (\scLatinEightCountSci).
To our knowledge, \tool is the first to obtain an exact model count without enumeration as a theorem about a term of a proof assistant.

\begin{table}[t]
\centering
\caption{Checkers for certified model counting and related checkers.}
\label{tab:neighbours}
\footnotesize
\setlength{\tabcolsep}{4pt}
\begin{tabular}{@{}>{\raggedright}p{2.2cm}>{\raggedright}p{2.3cm}>{\raggedright}p{4.45cm}>{\raggedright\arraybackslash}p{2.4cm}@{}}
\toprule
Checker & A run ends in (about) & Verified statement & Trusted beyond the logic\\
\midrule
\tool & a Lean theorem (a CNF term; the objects) & $\#_V(F)=N$ for the term~$F$; with an encoding lemma, the number of objects & Lean's compiler and runtime\\
cpog-lean-check~\cite{BryantNAH25} & an output count (a parsed file) & the checker's final state implies that the POG is equivalent to the parsed formula and has the computed count & Lean's compiler and runtime; two unproved lemmas (\cref{sec:trust})\\
cpog-check~\cite{BryantNAH25} & an output count (a parsed file) & none & the C program\\
SCPOG's checker~\cite{BryantTH25} & an output count (a parsed file) & the formula and the circuit agree on the variables that are not projected away, verified down to machine code & the hardware model of CakeML\\
VeriPB with CakePB~\cite{McCreeshNOT26} & a HOL4 theorem (the objects) & the number of solutions, obtained by enumerating them & the hardware model of CakeML\\
\bottomrule
\end{tabular}
\end{table}

cake\_lpr~\cite{TanHM21} is verified down to machine code and returns a verdict about a parsed file; Subercaseaux et al.~\cite{SubercaseauxNGC24} record this verdict as an axiom about a Lean term for the empty hexagon number.
Lean's \lean{bv_decide}~\cite{Boving25} proves bit-vector goals by checking the SAT solver's LRAT refutation, and lean-smt~\cite{MohamedMKBRQTB25} reconstructs SMT proofs as native Lean proofs.
lrat-catcher~\cite{Szeider26a} streams an LRAT refutation through Lean's checker; this way, even refutations of many gigabytes become Lean theorems.
Gallicchio et al.~\cite{GallicchioCAH26} obtain Lean theorems for Keller's conjecture up to dimension five using proof by reflection, that is, by running a checker inside Lean whose correctness is proved, and Kirchweger et al.~\cite{KirchwegerMS26} use Lean's LRAT checker for verified graph search with SAT Modulo Symmetries.
All of this work is based on checking refutations.

Sequencelib~\cite{MoreiraS26} is a Lean library for more than 25,000 OEIS sequences, containing the first values proved by computation.
With \tool, we prove single large values of such sequences. Dan\v{c}o et~al.~\cite{DancoJCA25} count finite algebraic structures up to isomorphism as the models of a formula with complete symmetry breaking. With Burnside's lemma, we count up to symmetry without symmetry breaking, as a sum of counts each of which is a theorem.

\section{Background}\label{sec:background}

\paragraph{Formulas and counts.}
We fix a \emph{universe} of $V$ propositional variables. A CNF formula over the universe is a set of clauses, which Lean and DIMACS files give as a list.
The \emph{model count} $\#_V(F)$ of a formula~$F$ is the number of assignments of the~$V$ variables that satisfy every clause of~$F$~\cite{Valiant79,GomesSS21}.
A variable in no clause doubles the count, so the universe is part of the statement; for a benchmark it is the number of variables declared in the DIMACS header.
In Lean, the formula~$F$ has the type \lean{CNF Nat} of the standard library.
Its variables are numbered from~$0$.
The universe consists of the variables $0,\dots,V-1$.
An assignment gives a truth value to every nonnegative integer, not only to the variables of the universe.

\paragraph{Partitioned-operation graphs.}
The trace of an exhaustive DPLL search with component caching is a circuit in \emph{decision-DNNF}~\cite{HuangD07,DarwicheM02}.
The exact counter d4~\cite{LagniezM17} can write this circuit to a file.
Other counters, such as ganak~\cite{SharmaRSM19}, output only the count; SharpSAT-TD~\cite{KorhonenJ21} has been extended to write the circuit~\cite{KieselE23}.
The compiler Panini~\cite{LaiMY25} writes circuits in the related language CCDD, and tree decision diagrams~\cite{CapelliCMMB26} are a recent canonical language between OBDDs and structured d-DNNF.
Bryant et al.~\cite{BryantNAH25} represent such circuits as \emph{partitioned-operation graphs} (POGs).
Each inner node is labeled with a fresh variable, its \emph{node variable}.
Our checker reads the POG from the node table (\cref{sec:contract}).
The $i$th node is the variable $V+i$, so the node variables are fresh by construction.
The arguments of an inner node are its out-neighbors.
The \emph{support} $S(u)$ of a node~$u$ is the set of input variables below it in the circuit, i.e., for a literal it is its variable and for an inner node it is the union of the supports of its arguments.
A POG is \emph{valid} if the arguments of every product have pairwise disjoint supports and the two arguments of every sum are not true together.
For a node~$u$, let $c(u)$ be the number of assignments of~$S(u)$ that make~$u$ true.
In a valid POG, a literal~$u$ has $c(u)=1$, a product with the set~$A$ of arguments has $c(u)=\prod_{a\in A} c(a)$, and the sum of $v$ and~$w$ has $c(u)=c(v)\cdot 2^{\,|S(u)|-|S(v)|}+c(w)\cdot 2^{\,|S(u)|-|S(w)|}$.
The root~$r$ of the circuit is true under $c(r)\cdot 2^{\,V-|S(r)|}$ assignments of the universe, which is $\#_V(P)$ in the chain~\eqref{eq:chain}.

\begin{figure}[t]
\centering
\begin{tikzpicture}[
  font=\footnotesize,
  gate/.style={draw, circle, inner sep=0pt, minimum size=5mm},
  leaf/.style={inner sep=1.5pt},
  cnt/.style={font=\scriptsize, inner sep=1pt, text=black!70},
  ln/.style={anchor=west, font=\ttfamily\footnotesize, inner sep=0pt},
  gone/.style={text=black!45},
  br/.style={decorate, decoration={brace, amplitude=3pt}}]
\node[gate] (s) at (1.5,2.3) {$\lor$};
\node[gate] (p) at (0.5,1.25) {$\land$};
\node[gate] (q) at (2.5,1.25) {$\land$};
\node[leaf] (a) at (0,0.25) {$\lnot x_1$};
\node[leaf] (b) at (1,0.25) {$x_2$};
\node[leaf] (c) at (2,0.25) {$x_1$};
\node[leaf] (d) at (3,0.25) {$x_3$};
\draw (s) -- (p) (s) -- (q) (p) -- (a) (p) -- (b) (q) -- (c) (q) -- (d);
\node[cnt, left=1pt of p] {$p_4$};
\node[cnt, right=1pt of q] {$p_5$};
\node[cnt, left=1pt of s] {$s_6$};
\node[cnt, right=2pt of s] {$c=1\cdot 2+1\cdot 2=4$};
\node[cnt] at (0.5,-0.25) {$c=1$};
\node[cnt] at (2.5,-0.25) {$c=1$};
\node[anchor=west, inner sep=0pt] at (-0.45,3.15) {$F=(x_1\lor x_2)\land(\lnot x_1\lor x_3)$};
\foreach \t/\y in {{r 6}/0, {3 p 4 -1 2 0}/1, {6 p 5 1 3 0}/2, {9 s 6 4 5 4 7 0}/3,
                   {12 a 4 1 0 1 3 0}/4, {13 a 5 -1 0 2 6 0}/5, {14 a 1 6 0 10 12 0}/6,
                   {15 a 6 0 14 11 13 0}/7}
  \node[ln] (l\y) at (4.95,3.15-0.31*\y) {\t};
\foreach \t/\y in {{d 14 10 12 0}/8, {d 13 2 6 0}/9, {d 12 1 3 0}/10,
                   {d 1 15 5 7 9 0}/11, {d 2 15 4 8 9 0}/12}
  \node[ln, gone] (l\y) at (4.95,3.15-0.31*\y) {\t};
\draw[br] (8.3,3.30) -- (8.3,2.08) node[midway, right=4pt, align=left] {root and declarations\\(validity check)};
\draw[br] (8.3,2.02) -- (8.3,0.84) node[midway, right=4pt, align=left] {additions\\(forward check)};
\draw[br, black!45] (8.3,0.78) -- (8.3,-0.70) node[midway, right=4pt, align=left, text=black!55] {deletions,\\not generated};
\end{tikzpicture}
 \caption{A formula~$F$, a POG for it with the counts of its nodes, and its CPOG certificate. The clauses of~$F$ have the identifiers 1 and~2; the declaration \texttt{3 p 4 -1 2 0} introduces the product node~4 of $\lnot x_1$ and~$x_2$ with the defining clauses 3 to~5, \texttt{s} declares a sum (the two numbers before its \texttt{0} are the exclusion hint), \texttt{r 6} names the root, and an addition \texttt{a} gives the identifier, the clause and, after the first~\texttt{0}, its hints.}
\label{fig:certificate}
\end{figure}
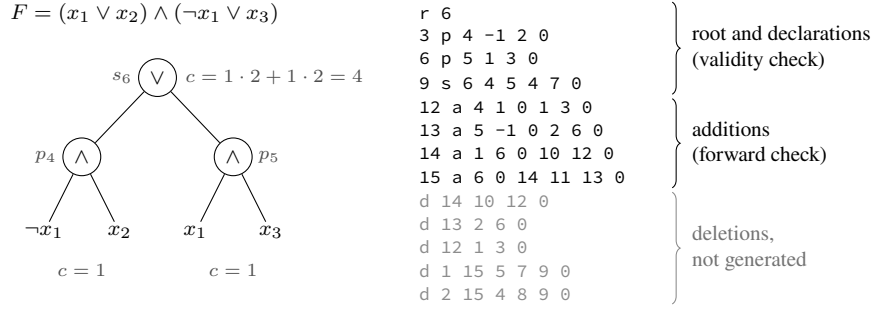
 
\paragraph{CPOG certificates.}
A CPOG certificate~\cite{BryantNAH23,BryantNAH25} is a proof that a POG with root~$r$ is equivalent to the formula~$F$.
Its lines form three groups indicated in \cref{fig:certificate}: (i) the declarations of the nodes and the root, (ii) the additions, and (iii) the deletions.
The declarations mention the nodes one by one, each with its Tseitin \emph{defining clauses}, which are the clauses that make the node variable equivalent to the product or the sum of its arguments.
A \emph{hint} is a list of earlier clauses.
If we assume the negation of a new clause and apply \emph{unit propagation} to the clauses of the hint, then the new clause follows if a conflict is reached.
This is exactly the reverse unit propagation (RUP) criterion of LRAT~\cite{Cruz-FilipeHHKS17}.
The declaration of a sum has an \emph{exclusion hint}: defining clauses from which the checker derives by unit propagation that the two arguments exclude each other.
The additions are a proof of the forward implication ($F$ with the defining clauses implies the root).
Each added clause has a hint from which it follows by RUP.
The last addition is the unit clause of the root.
The deletions are a proof of the reverse implication (the circuit implies every clause of~$F$).

\paragraph{Importing refutations into Lean.}
An LRAT \emph{refutation} of a formula is a sequence of additions with hints that ends with the empty clause.
lrat-catcher~\cite{Szeider26a} turns such a refutation into a Lean theorem (a CNF formula given as a Lean term is unsatisfiable).
It runs the verified LRAT checker of Lean's standard library~\cite{Boving25} on the refutation and obtains the theorem by means of the checker's soundness lemma, based on native evaluation (\cref{sec:intro,sec:trust}).
Refutations that are too large to be embedded in the Lean file are streamed, i.e., read in blocks during the evaluation.

\section{From Certificate to Theorem}\label{sec:design}

The chain~\eqref{eq:chain} from the number of objects to~$N$ consists of four equalities.
The first is the encoding lemma, which the user provides (\cref{sec:lemmas}).
In this section, we prove the other three: the equivalence of formula and POG (forward and reverse check), the count of the POG (validity check) and the arithmetic that gives~$N$.

\subsection{The Theorem and the Contract}\label{sec:contract}

In the count theorem, the formula~$F$ is the Lean term of type \lean{CNF Nat}.
An \emph{encoder} is a Lean function that maps a parameter, such as the order of the Latin squares, to a formula~$F$.
Usually the formula~$F$ is an encoder applied to a parameter, for example, the encoder for domino tilings applied to a board.
The certificate occurs only in the hypothesis \lean{h} of \lean{checkAll_sound} (for an embedded certificate, as the strings \lean{pogStr} and \lean{lratStr}), and not in the count theorem.
For a wrong count~$N$, the checker rejects and there is no theorem.
No file and no parser occur in the count theorem, with one exception.
For a formula that exists only as a DIMACS file, $F$ is a parser written in Lean applied to the text of the file.
The proof does not depend on the correctness of the parser.
However, we have to trust the parser when we interpret the theorem as a statement about the benchmark instance in the file.

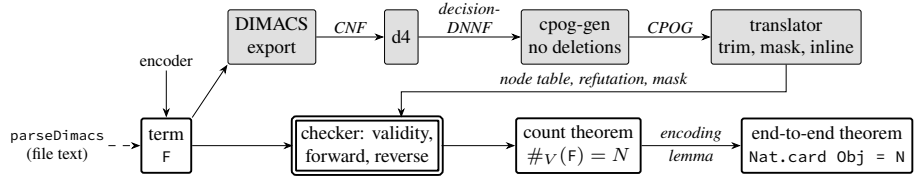
\begin{figure}[t]
\centering
\scalebox{0.9}{%
\begin{tikzpicture}[
  font=\fontsize{8}{9.5}\selectfont, >={Stealth[length=4pt]},
  adv/.style={draw, fill=black!12, rounded corners=1pt, inner sep=3pt, align=center, minimum height=8mm},
  lean/.style={draw, thick, rounded corners=1pt, inner sep=3pt, align=center, minimum height=8mm},
  nat/.style={draw, thick, double, double distance=1pt, rounded corners=1pt, inner sep=3pt, align=center, minimum height=8mm},
  dat/.style={inner sep=1.5pt, align=center, font=\fontsize{7}{8}\selectfont\itshape},
  src/.style={inner sep=1.5pt, align=center, font=\fontsize{7}{8}\selectfont}]
\node[adv]  (exp) at (2.3,1.6) {DIMACS\\export};
\node[adv]  (d4)  at (4.2,1.6) {d4};
\node[adv]  (gen) at (6.75,1.6) {cpog-gen\\no deletions};
\node[adv]  (tr)  at (9.85,1.6) {translator\\trim, mask, inline};
\node[lean] (F)   at (0.75,0) {term\\\texttt{F}};
\node[nat]  (chk) at (3.7,0) {checker: validity,\\forward, reverse};
\node[lean] (thm) at (6.8,0) {count theorem\\$\#_V(\texttt{F})=N$};
\node[lean] (obj) at (10.45,0) {end-to-end theorem\\\texttt{Nat.card Obj = N}};
\node[src] (s1) at (0.75,1.25) {encoder};
\node[src] (s2) at (-0.85,0) {\texttt{parseDimacs}\\(file text)};
\draw[->] (s1) -- (F);
\draw[->, dashed] (s2) -- (F);
\draw[->] (F.north east) -- (exp.south west);
\draw[->] (exp) -- node[dat, above] {CNF} (d4);
\draw[->] (d4) -- node[dat, above] {decision-\\DNNF} (gen);
\draw[->] (gen) -- node[dat, above] {CPOG} (tr);
\draw[->] (tr.south) -- ++(0,-0.4) -| node[dat, above, pos=0.25] {node table, refutation, mask} ([xshift=5mm]chk.north);
\draw[->] (F) -- (chk);
\draw[->] (chk) -- (thm);
\draw[->] (thm) -- node[dat, above] {encoding} node[dat, below] {lemma} (obj);
\end{tikzpicture}

}
\caption{From a term to a theorem. Shaded components are untrusted; thick frames are Lean terms and theorems, the double frame the checker, proved sound in Lean and run as native code; the dashed arrow is the path of a formula given as a DIMACS file.}
\label{fig:pipeline}
\end{figure}

We show the origin of the certificate in \cref{fig:pipeline}.
\tool exports the term to DIMACS in its clause order, so that clause identifiers agree.
d4 compiles the formula into a decision-DNNF, and cpog-gen generates the certificate; we patched cpog-gen to omit the deletions that prove the reverse implication.
The translator of \tool is an untrusted program that splits the certificate into a node table (containing the declarations) and an LRAT refutation.
The translator may shorten the refutation (\cref{sec:partb}).
The checker starts from the term~$F$, not from a file.
Consequently, a failure in any of these programs can cause the check to fail, but it cannot lead to a false theorem.

\begin{table}[t]
\centering
\caption{The contract of the check. $\Phi$ consists of the clauses of~$F$, the defining clauses of~$P$ and the unit clause~$\lnot r$; keep mask: \cref{sec:partb}, implication masks: \cref{sec:partc}.}
\label{tab:contract}
\footnotesize
\setlength{\tabcolsep}{4pt}
\begin{tabular}{@{}l>{\raggedright}p{2.45cm}>{\raggedright}p{2.45cm}>{\raggedright}p{2.55cm}>{\raggedright\arraybackslash}p{2.55cm}@{}}
\toprule
Check & Establishes & Built in Lean & Supplied, untrusted & Checked by\\
\midrule
validity & $P$ is a valid POG with count~$c(r)$ & $P$ and~$r$, parsed from the node table & node table: declarations, exclusion hints & one pass over the node table\\
forward & $F$ implies the root ($\Phi$ is unsatisfiable) & $\Phi$, from~$F$ and~$P$ & additions with hints, keep mask & Lean's LRAT checker, run by lrat-catcher\\
reverse & the root implies~$F$ & implication masks, from~$P$ and~$F$ & none & bit-parallel evaluation\\
\bottomrule
\end{tabular}
\end{table}

\Cref{tab:contract} states the contract: for each check, what Lean builds itself and what the certificate supplies.
All three checks refer to the same POG~$P$ with root~$r$, which is parsed once from the node table.
We combine them into the chain~\eqref{eq:chain} with the composition theorem \lean{card_models_eq_countRoot}.
The checker verifies that every variable~$x$ of~$F$ satisfies $x<V$ and that the node table states the same number~$V$ of variables.
It also compares $c(r)\cdot 2^{\,V-|S(r)|}$ with~$N$.
A small certificate can be embedded in the Lean file and checked by a single native evaluation.
For a streamed refutation (\cref{sec:background}), we prove the forward implication as a separate theorem (\cref{lst:soundness} in the appendix).

\subsection{The Validity Check}\label{sec:parta}

With the validity check, we establish that the POG of the node table is valid and represents the computed count (lemma \lean{checkPog_sound}).
In one pass the checker verifies that (i) every node refers only to the $V$ input variables and to earlier nodes, (ii) the arguments of a product have pairwise disjoint supports, and (iii) the two arguments of a sum exclude each other, and it computes the support and the count of every node.

As in CPOG~\cite{BryantNAH25}, an exclusion hint may reference defining clauses only, and the parser rejects a reference not pointing to a defining clause.

The checker accepts the fragment of CPOG that cpog-gen produces: (i) a sum has two arguments, (ii) a product any number of arguments, (iii) no argument is a negated node, and (iv) the root is a literal of an input variable or an unnegated node. The reverse check tests (iii) and (iv) (\cref{app:adversarial}).
The constant false is not in the fragment, and cpog-gen does not produce certificates for unsatisfiable formulas.
Therefore, we prove a count of zero from an LRAT refutation of~$F$.
We do not preprocess.
Consequently, variables that appear in no clause stay in the universe, and duplicate clauses, tautologies and repeated literals stay in the formula.

\subsection{The Forward Check}\label{sec:partb}

Let $\Phi$ be the set of all clauses of~$F$, the defining clauses of~$P$ and the unit clause~$\lnot r$.
The forward implication is true if and only if $\Phi$ is unsatisfiable (lemma \lean{fwd_of_unsat}).
The additions of the certificate, followed by the empty clause, form an LRAT refutation of~$\Phi$.

Lean generates~$\Phi$ itself from~$F$ and the parsed node table.
From the refutation file, Lean only reads the additions and their hints.
Consequently, the checker accepts a refutation only if it refutes~$\Phi$, whatever formula, POG or defining clauses the translator used.

Using the same reasoning, the translator can shorten the refutation.
It proceeds in three steps.
cpog-gen adds a clause for every unit that it propagates.
However, our translator keeps only those additions on which the empty clause depends.
We call this trimming.
For a streamed refutation, it saves the kept additions in the binary LRAT format.
It also saves a \emph{keep mask} with one bit set for each clause of~$\Phi$ that a kept addition cites.
The checker loads only the marked clauses.
A refutation of a subset of~$\Phi$ is also a refutation of~$\Phi$.
Lean selects this subset from~$\Phi$, which it generated itself.
Consequently, the keep mask can only exclude clauses but cannot add clauses.
Finally, an addition that is cited only once is inlined, i.e., its hints are copied into the step that cites it, which the translator does only where they propagate there under the checker's rule.

\subsection{The Reverse Check}\label{sec:partc}

SCPOG omits the deletions (\cref{sec:background,sec:related}), and its checker evaluates the POG once per clause~\cite{BryantTH25}.
We check a batch of 63 clauses at once.
For every node, we propagate the set of clauses of the batch that the node implies as the bits of a number below~$2^{63}$, its \emph{implication mask}, which Lean's runtime keeps in one machine word.

A node \emph{implies} the clause~$D$ if every assignment of the input variables that makes the node true satisfies~$D$.
For the fragment that the checker accepts (\cref{sec:parta}), we can decide whether a node implies~$D$ from its arguments in a bottom-up process.
A literal of an input variable implies~$D$ if it occurs in~$D$.
A product implies~$D$ if at least one of its arguments does, and a sum if both of its arguments do.
If the root implies every clause of~$F$, then it implies~$F$, which is the reverse implication.
The checker skips tautological clauses and it stores an implication mask for every node.
It checks the whole batch in one pass over the nodes with one word operation per edge (\cref{alg:partc} in the appendix).
In \cref{fig:certificate}, the products~4 and~5 imply both clauses of~$F$, node~4 through~$x_2$ and~$\lnot x_1$ and node~5 through~$x_1$ and~$x_3$, so the sum~6 implies both and the check accepts.
Node~4 would imply a clause~$(x_2)$ but node~5 would not, so the sum would not either, and the check would reject.

We prove in lemma \lean{checkRevKillFast_sound} that acceptance implies the reverse implication.
The proof is based on the invariant that bit~$j$ of the mask of a literal or node is set only if it implies the clause~$D_j$ of the batch.
We show this by induction over the node table.
This is the only implication required for soundness.
In the fragment that the checker accepts, the converse is also true: bit~$j$ is set whenever the node implies~$D_j$.
Consequently, if the root implies~$F$, the reverse check accepts.

\subsection{Design Rationale and Trusted Base}\label{sec:trust}

Running CPOG's verified checker~\cite{BryantNAH25} inside Lean by reflection would not give the count theorem directly, for two reasons.
First, its correctness theorem is about the checker's final state and the parsed file, not about the Lean term (\cref{sec:intro}).
Second, it counts every node over all declared variables~\cite{BryantNAH25,Nawrocki26}.
So we would still need to prove that its result is the model count of the formula~$F$ over the universe.
Therefore, we created a new validity check and reverse check, with a new composition theorem (\cref{sec:contract}).

The theorem of \tool depends on the Lean kernel, the three standard axioms, and one axiom per native evaluation: one for an embedded certificate, two for a streamed one.
Each native evaluation adds an axiom stating that the evaluated Boolean term is true, for example, \lean{checkAll V F pogStr lratStr N = true}.
For a streamed refutation, the term contains a constant of type \lean{Nat → Option (Array IntAction)}.
It has no definition in the logic.
Only its native implementation runs during the evaluation.
For a given~$i$, it returns the $i$th block of additions of the refutation file.
The soundness lemma is true for every function of this type.
So, in particular, the file is not trusted.
\lean{count_models} proves the forward implication of a streamed refutation as a separate theorem about the same formula~$F$ and the same POG constant, and \lean{checkAllAC_sound} (\cref{lst:soundness}) takes it as a hypothesis.
Our trusted base contains the Lean compiler and runtime (with GMP for large numbers), including the unproved substitutions of library functions by faster ones, and the agreement of the precompiled checker with the definitions that the kernel checks, which Lean does not test at load time (\cref{app:soundness}).
This is the same trusted base as used by \lean{bv_decide}~\cite{Boving25}.
For a streamed refutation, each theorem has its own constant, so its axiom fixes only the blocks that arrived, and we trust that the native code returns every block exactly once.
Mathlib does not admit native evaluation, so our theorems cannot enter Mathlib as they stand, although they use its definitions.

Our proofs establish statements about the Lean definitions of models and model counts.
A mistake in these definitions would not show up in the proofs.
Therefore, we also test the checker.
For every condition of the three checks, we constructed certificates that violate exactly this one.
The checker rejected them all.
We also ran a differential test on small formulas that compares the checker's verdict with the count obtained by enumeration.
Of more than twelve million generated and mutated certificates, the checker accepted only those with a correct count.

We carried out a comparison between our trusted base and that of cpog-lean-check.
We observed that its archived build depends on two unproved lemmas of a pinned library.
We gave the two lemmas of the checker that use them direct proofs, which removes this dependency and leaves the checker's executable unchanged.
We sent these proofs to the authors of cpog-lean-check.

\section{Counts as Lemmas}\label{sec:lemmas}

\subsection{Encoding Lemmas}\label{sec:families}

With \emph{family}, we refer to an encoder (\cref{sec:contract}) together with the type~$X$ of objects, which is defined without a reference to the encoding, and an encoding lemma.
The encoding lemma is a Lean lemma that gives a bijection between the models of~$F$ over the universe and~$X$.
With the lemma \lean{card_of_check} (\lean{card_of_checkAC} for a streamed refutation), shared by all families, we obtain \lean{Nat.card X = N} from $\#_V(F)=N$.

Our encoders use no auxiliary variables.

\begin{lstlisting}[float=t,caption={The Latin squares of order~7: the first of the four \lean{count_models} lines and the reduced count (\cref{sec:burnside}; proof term abbreviated, full file in the artifact); \texttt{fwd.pog} is the node table, \texttt{fwd.lrat} the refutation and \texttt{fwd.keep} the keep mask; the last comment summarizes \texttt{\#print axioms}, two native evaluations per streamed fiber.},label=lst:latin]
count_models latin7_fibre0
  via (Latin.fibre_card_of_check (n := 7) (r := ![1, 0, 3, 2, 5, 6, 4]))
  "c0/fwd.pog" "c0/fwd.lrat" (keep := "c0/fwd.keep") (count := 55296)
theorem latin7_reduced :
    Nat.card {L : Fin 7 → Fin 7 → Fin 7 // IsReducedLatin L}
      = 16942080 :=
  Latin.card_eq_of_classes (by norm_num) 16942080
    [(![1, 0, 3, 2, 5, 6, 4], 55296), ...]
    (List.forall_mem_cons.2 ⟨latin7_fibre0, ...⟩) ...
    (by decide +kernel) (by decide +kernel)
-- propext, Classical.choice, Quot.sound, and 8 native evaluations
\end{lstlisting}
 
\Cref{lst:latin} shows the theorem file of the Latin squares of order~7, the order-7 counterpart of \lean{latin8_all} in \cref{sec:intro}.
We contribute the encoder and the encoding lemma.
The encoder uses one variable per cell and symbol, exactly-one clauses per cell and, for every symbol, per row and per column, and unit clauses to fix the first row and first column.
As guaranteed by the encoding lemma, the models are in bijection with the reduced Latin squares of order~7.
We split the reduced squares by their second row into \emph{fibers} (\cref{sec:burnside}).
For the first fiber, \tool is called as
\begin{lstlisting}
count-catcher latin7_fibre0 \
  --via 'Latin.fibre_card_of_check (n := 7) (r := ![1, 0, 3, 2, 5, 6, 4])'
\end{lstlisting}
This call, one per fiber, exports the formula with \lean{#export_cnf}, runs the pipeline of \cref{fig:pipeline}, and writes the \lean{count_models} line that opens the listing.
A Python script of \tool then appends the theorem \lean{latin7_reduced}, which puts the fiber counts together (\cref{sec:burnside}).
No line of the listing is written by hand.
If a check fails, \lean{count_models} reports which check failed and where: the node and the condition for the validity check, the LRAT step for the forward check, and for the reverse check a clause of~$F$ that the root does not imply, with an assignment that satisfies the root and falsifies the clause.
The encoders do not occur in the theorem statements.
The only definitions there not covered by Mathlib are \lean{IsLatin} (\cref{sec:intro}) and its reduced variant.

For a new family, one is required to write the encoder and the encoding lemma.
For instance, these are \famKrewerasLines{} lines of Lean for the antichains of the OEIS sequence A358391 (\cref{sec:eval}) and \famLatinLines{} for the Latin squares with the lemmas of their fiber split.
For the independent sets and perfect matchings of a graph (where the domino tilings of a board are the perfect matchings of its grid graph), the encoder and encoding lemma are generic and work for every graph.
For a new instance, we only need to supply the graph.

\subsection{Composition by Symmetry}\label{sec:burnside}

There are \scTriTenCount{} triangle-free graphs on ten vertices up to isomorphism~\cite{OEIS-A006785}.
Counting up to isomorphism by Burnside's lemma over the cycle types is a standard technique~\cite{Polya37} (see~\cite[Chapter~2]{HararyP73}).
In our Lean theorem that counts the triangle-free graphs on ten vertices up to isomorphism, every summand is itself a theorem, namely one of the 42 certified counts of \cref{sec:intro}.

Mathlib has no action of vertex permutations on \lean{SimpleGraph}.
We define such an action by relabeling the $n$ vertices, restricted to graphs with an invariant property such as being triangle-free.
We prove in Lean that the orbits of this action are exactly the isomorphism classes.
Consequently, the number of isomorphism classes equals the number of orbits, which Burnside's lemma computes.
Two permutations with the same cycle type fix the same number of graphs.
Therefore, we can regroup Mathlib's Burnside sum into one summand per cycle type.
Each summand is the number of permutations with the cycle type, which is the class size, times the number of graphs that one of them fixes.
However, the list of cycle types in the proof is an untrusted input, like a certificate.
Using Lean's \lean{decide}, we verify that (i) the entries are cycle types of $n$ elements and pairwise distinct, (ii) the class sizes add up to~$n!$, and (iii) the weighted sum is the claimed number of orbits times~$n!$.

We get the second factor of each summand as a model count, since the permutation of the vertices induces a permutation~$\pi$ of the edge variables.
The graphs that $\pi$ fixes are the models of the encoding~$F$ extended by the two binary clauses for $x\leftrightarrow\pi(x)$, for every variable~$x$ that $\pi$ moves.
We prove this once for any group acting on the variables and on the objects such that the bijection of the encoding lemma commutes with both actions.
A separate proof that $\pi$ preserves~$F$ is not needed.
Consequently, each fixed-point count is a separate theorem, where the encoding lemma is instantiated by the cycle type.

For the Latin squares of \cref{lst:latin}, we group the count of the reduced squares by cycle type in the same way.
The number of all squares is $n!\,(n-1)!$ times the number of reduced ones.
We establish this equality in Lean.
The fiber of a second row consists of the reduced squares with that row.
This is encoded by the encoder plus one unit clause per cell of the row.
Applying the same permutation~$\sigma$ to the column indices and to the symbols keeps the first row the identity and turns a second row~$r$ into $\sigma r\sigma^{-1}$, which has the same cycle type.
If $\sigma r\sigma^{-1}$ is again a second row, a permutation of the other rows makes the first column the identity again, so the fiber of~$r$ and the fiber of~$\sigma r\sigma^{-1}$ have the same size.
Consequently, the fiber count depends only on the cycle type.
For order~8, unit clauses on three cells of the third row split each fiber further, and we prove in Lean that the parts add up to the fiber.

\section{Evaluation}\label{sec:eval}

\begin{table}[t]
\centering
\caption{Theorems about mathematical objects: ``Summands'' is the number of certified counts in the proof, ``Count'' the number of objects (for the graphs, of isomorphism classes); for the Latin squares, ``Lean'' is summed over the certificates and the assembled theorem. $^\dagger$: not listed in the OEIS at time of writing.}
\label{tab:showcases}
\footnotesize
\setlength{\tabcolsep}{4pt}
\begin{tabular}{@{}>{\raggedright\arraybackslash}p{7.2cm}rrr@{}}
\toprule
Theorem & Summands & Count & Lean (s)\\
\midrule
\multicolumn{4}{@{}l}{\emph{Composition of certified counts}} \\
\quad Triangle-free graphs on 10 vertices, A006785 & \scTriTenComposed & \scTriTenCount & \scTriTenLean \\
\quad Latin squares of order 7 & \scLatinSevenComposed & \scLatinSevenCountSci & \scLatinSevenLean \\
\quad Latin squares of order 8 & \scLatinEightComposed & \scLatinEightCountSci & \scLatinEightLean \\
\addlinespace
\multicolumn{4}{@{}l}{\emph{One count as an end-to-end theorem}} \\
\quad Dedekind's number $M(7)$ & \scDedekindSevenComposed & \scDedekindSevenCountSci & \scDedekindSevenLean \\
\quad A358391(6), Kreweras$^\dagger$ & \scKrewerasSixComposed & \scKrewerasSixCountSci & \scKrewerasSixLean \\
\quad A396728(7), Sperner$^\dagger$ & \scSpernerSevenComposed & \scSpernerSevenCountSci & \scSpernerSevenLean \\
\quad A303210(9), total domination$^\dagger$ & \scTotDomNineComposed & \scTotDomNineCountSci & \scTotDomNineLean \\
\quad A303210(10), total domination$^\dagger$ & \scTotDomTenComposed & \scTotDomTenCountSci & \scTotDomTenLean \\
\addlinespace
\multicolumn{4}{@{}l}{\emph{A count as a hypothesis}} \\
\quad The $(3,5)$-Ramsey graph on 13 vertices is unique & \scRamseyThirteenComposed & \scRamseyThirteenCount & \scRamseyThirteenLean \\
\bottomrule
\end{tabular}
\end{table}

For \cref{tab:showcases}, we chose well-known problems that need counting rather than enumeration.
Dedekind's number~$M(8)$ was out of reach for d4 in our probes, and for Latin squares of order~7, the generation of one certificate for all reduced squares exceeded 64~GB, so we split by fibers.

\paragraph{Composition of certified counts.}
The first group of \cref{tab:showcases} consists of the compositions of \cref{sec:burnside}; the largest is \lean{latin8_all} of \cref{sec:intro}, assembled from \scLatinEightLeaves{} certified counts in \scLatinEightLeanHours{} hours of Lean time, and we checked the count against the literature~\cite{Wells67}.

\paragraph{One count as an end-to-end theorem.}
We state Dedekind's number $M(7)$ with Mathlib's \lean{Monotone} and no definition of our own.
The four OEIS values of \cref{sec:intro} are A358391(6)~\cite{OEIS-A358391}, the antichains of the lattice of noncrossing partitions of a six-element set; A396728(7)~\cite{OEIS-A396728}, the Sperner families on seven elements that contain no set with its complement; and A303210(9) and A303210(10)~\cite{OEIS-A303210}, the total dominating sets of the torus grid.
For even~$n$ the torus is bipartite and every clause lies within one side, so the formula splits into two independent halves, which makes A303210(10) cheaper than A303210(9).
As a check outside our pipeline, which uses d4, we ran the counters ganak~\cite{SharmaRSM19} and SharpSAT-TD~\cite{KorhonenJ21} on the same formulas, and both give the same four integers.

\paragraph{A count as a hypothesis.}
The Ramsey graph of \cref{sec:intro} is unique, a classical fact~\cite{Kery64,Radziszowski}.
Using certified counts, we show that there is exactly one isomorphism class.
The labeled count, the summand of the identity, is itself split by the neighborhood of one vertex into 13 certified counts, hence \scRamseyThirteenComposed{} summands for the 101 cycle types of 13 elements.
The count of one is the hypothesis of the uniqueness theorem: with a separate check, we show that Mathlib's circulant graph with jumps 1 and~5 belongs to this class, so every such graph is isomorphic to it.

\paragraph{Setup.}
For all tables and figures of this section, we ran the pipeline of \cref{fig:pipeline} with the defaults of \cref{sec:design} (no deletions, shortened refutation), except where a row or \cref{tab:app-provenance} states otherwise.
The jobs had access to two cores on shared machines of a Slurm cluster; the appendix lists the sources, tools, limits and repetitions of every table and figure (\cref{app:provenance}).
We compare with CPOG's checkers, running both in the same job.
Since absolute times vary by tens of percent between runs on shared machines, we compare ratios and, when summarizing, medians.
\emph{Lean} refers to the whole Lean process, including the import of Mathlib and the processing of the theorem file.
\emph{Check} is the time of the three checks as \lean{count_models} reports it.
\emph{Memory} is the largest amount of main memory that the Lean process occupied, of which at most \cmpOurMemMin~GB are Mathlib's compiled files, which Lean loads into memory.
We used the public instances of the unweighted and the weighted track of the 2022 Model Counting Competition~\cite{FichteH23,FichteH26}, \funPerTrack{} formulas each. We ignore the weights and count models. Without its weights, each of \funDuplicates{} formulas of the weighted track equals a formula of the unweighted track, so \funTotal{} formulas remain. Bryant et~al.~\cite{BryantNAH25} evaluate CPOG on the same \funTotal{} formulas, which allows a direct comparison with their results.
We refer to this set of formulas as the \emph{competition set}.
For the \famN{} instances of four validation families (domino tilings, queens, independent sets of grids and of graphs), we prove the counts in Lean.
All these counts agree with the OEIS or with two other counters (\cref{app:families}).

\paragraph{Results on the competition set.}
In the appendix (\cref{tab:funnel}), we follow the \funTotal{} formulas through the four stages of the pipeline, and \funProved{} become theorems.
Of the formulas that do not become theorems, all but one fail in d4 or the generator.
Our checker rejects no certificate it receives, but times out on one certificate of \funLeanTimeoutCertGB~GB.
The generator cpog-gen ran with our two patches (\cref{app:provenance}) and the same time limit of \funLimitGen~s as d4.
It derived the refutation from the structure of the POG.
With cpog-gen and the unverified checker cpog-check, Bryant et al.~\cite{BryantNAH25} verify 111 of the 123 formulas of the set that d4 solves within \funLimitDFour~s.
Their generator switches to one SAT call per subgraph below a tree-size threshold, and they allow 10,000~s for generation, check and count together.
In a rerun of the \rgFailN{} formulas on which our generator had failed, with a generation limit of 10,000~s, \rgThm{} more become theorems, \rgTotal{} in total (\cref{app:competition}).
Switching to SAT calls as in their generator adds none, and the remaining formulas exceed 10,000~s or 64~GB.
On the proved formulas the median Lean process takes \funLeanMed~s, about half of the \funGenMed~s of generation (d4: \funDFourMed~s); the largest certificate checked has \bigCertGB~GB, the largest count \bigCountDigits{} decimal digits.
These formulas exist only as DIMACS files, so their theorems are about the parsed term of \cref{sec:contract}, with a trusted base comparable to that of cpog-lean-check (\cref{sec:trust}), and serve only as a scaling test.

\begin{figure}[t]
\centering
\includegraphics{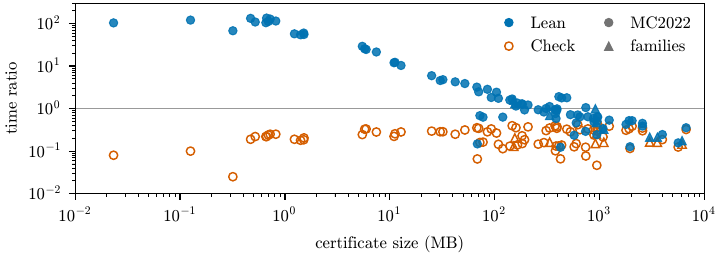}\\[6pt]
\footnotesize
\setlength{\tabcolsep}{3pt}
\begin{tabular}{@{}lrrrrrr@{}}
\toprule
Measured & \tool & Verified & Unverified & Ratio & Lower & $n$\\
\midrule
Checks after translation (s) & \cmpOurCheckMed & \cmpVcMed & \cmpCcMed & \cmpBoundCheckMed & \cmpBoundCheckLower & \cmpBoundCheckN \\
Translation and Lean run (s) & \cmpThmOursMed & \cmpVcMed & \cmpCcMed & \cmpBoundThmMed & \cmpBoundThmLower & \cmpBoundThmN \\
Pipeline after d4 (s) & \cmpPipeOursMed & \cmpPipeVcMed & \cmpPipeCcMed & \cmpBoundPipeMed & \cmpBoundPipeLower & \cmpBoundPipeN \\
Peak memory (GB) & \cmpOurMemMed & \cmpVcMemMed & \cmpCcMemMed & \cmpBoundMemMed & \cmpBoundMemLower & \cmpBoundMemN \\
Checks, unshortened refutation & -- & -- & -- & \unshCheckMed & \unshCheckLower & \unshN \\
Correct count reported & \cmpOursProved & \cmpVcCorrect & \cmpCcCorrect & -- & -- & \cmpN \\
\bottomrule
\end{tabular}
\caption{The cost of the theorem against CPOG's checkers on the same d4 outputs. \emph{Top}: the time of \tool over the time of the verified checker per instance, against the size of the full certificate; the line is ratio one. \emph{Bottom}: medians over the $n$ jointly completed runs; ``Ratio'' is the median per-instance ratio of \tool over the verified checker, ``Lower'' the number of instances on which \tool is lower.}
\label{fig:compare}
\end{figure}

\begin{table}[t]
\centering
\caption{Medians of the per-instance ratios of \tool over cpog-lean-check by the size of the full certificate; $n$ is the number of jointly completed runs.}
\label{tab:bands}
\footnotesize
\setlength{\tabcolsep}{5pt}
\begin{tabular}{@{}lrrrr@{}}
\toprule
Certificate size & $n$ & Lean & Check & Memory\\
\midrule
\cmpBandOneLabel & \cmpBandOneN & \cmpBandOneLeanRatio & \cmpBandOneCheckRatio & \cmpBandOneMemRatio \\
\cmpBandTwoLabel & \cmpBandTwoN & \cmpBandTwoLeanRatio & \cmpBandTwoCheckRatio & \cmpBandTwoMemRatio \\
\cmpBandThreeLabel & \cmpBandThreeN & \cmpBandThreeLeanRatio & \cmpBandThreeCheckRatio & \cmpBandThreeMemRatio \\
\cmpBandFourLabel & \cmpBandFourN & \cmpBandFourLeanRatio & \cmpBandFourCheckRatio & \cmpBandFourMemRatio \\
\bottomrule
\end{tabular}
\end{table}

\paragraph{Time and memory against CPOG's checkers.}
We compare with the two checkers of CPOG~\cite{BryantNAH25}, the verified cpog-lean-check and the unverified cpog-check in C, on the same d4 outputs.
The comparison uses \cmpNMcTwentyTwentyTwo{} formulas of the competition set, those proved with generation under \cmpRuleGenLimit~s, and \cmpNFamilies{} of the \famN{} validation instances (without the two largest).
The limit, under which the generator's second run for CPOG's checkers fits our time budget, excludes 14 proved formulas.
The checkers read the full certificate with its deletions, written by the unpatched generator; \tool reads the node table and the refutation that the translator writes from the certificate without deletions.
We also measure the total time after d4 has finished, including certificate generation on both pipelines.
In the median, the theorem pipeline takes about the same time as the checked-count pipeline (\cref{fig:compare}).

The runtime is dominated by Lean's start-up.
Without it, \emph{Check} takes about a quarter of cpog-lean-check's time in the median of every size range (\cref{tab:bands}) and is faster on every instance.
The forward check takes \cmpFwdSharePct\% of \emph{Check} in the median (\cmpFwdShareN{} runs; the validity and reverse checks are timed together).
Part of this speed-up is due to the shortened refutation (\cref{sec:partb}).
In a second run on the same d4 outputs without trimming, keep mask or inlining, \emph{Check} is still faster than cpog-lean-check on every instance (\cref{fig:compare}, bottom).
We attribute most of the rest to the bit-parallel reverse check, which decides the reverse implication by evaluation instead of checking the deletions (\cref{sec:partc}).
For the whole Lean process, the start-up dominates below 100~MB of certificate.
Above 1~GB, \tool takes about a third of cpog-lean-check's time (\cref{tab:bands}).

Most of the memory is taken by Mathlib.
Because of the streaming, the checker's memory does not grow with the refutation size.
cpog-lean-check, however, keeps the certificate in memory.
Above 1~GB of certificate, the Lean process uses about half of cpog-lean-check's memory, and below that size, it uses more (\cref{tab:bands}).
Both checkers accepted every certificate.
cpog-lean-check reported the correct count on every one, and cpog-check a wrong count on \cmpCcWrong{} of them (\cref{fig:compare}, last row).
The reason is a fault in the normalization of rational numbers (\cref{app:miscount}).

\begin{table}[t]
\centering
\caption{Ablation: the median and the range of the factor by which the metric changes without the design choice (above 1: worse); ``clauses of $\Phi$ loaded'' is the percentage of the clauses of~$\Phi$ that the checker loads with the keep mask. The generation-time factor includes our second patch, applied only to the generator without deletions. The top-down evaluation is our unverified Lean implementation of the strategy of SCPOG's checker, the sweep per clause our earlier verified reverse check, one bottom-up pass over the POG per clause.}
\label{tab:ablation}
\footnotesize
\setlength{\tabcolsep}{4pt}
\begin{tabular}{@{}llrrr@{}}
\toprule
Design choice & Metric & Instances & Median & Range\\
\midrule
Deletions left out & certificate bytes & \ablRevBytesN & \ablRevBytesMed & \ablRevBytesMin--\ablRevBytesMax \\
 & generation time & \ablRevGenN & \ablRevGenMed & \ablRevGenMin--\ablRevGenMax \\
Bit-parallel reverse check & vs.\ top-down evaluation & \ablPartcTdN & \ablPartcTdMed & \ablPartcTdMin--\ablPartcTdMax \\
 & vs.\ sweep per clause & \ablPartcPcN & \ablPartcPcMed & \ablPartcPcMin--\ablPartcPcMax \\
Trimmed binary refutation & forward check time & \ablTrimBinPartBN & \ablTrimBinPartBMed & \ablTrimBinPartBMin--\ablTrimBinPartBMax \\
Keep mask & Lean time & \ablKeepLeanN & \ablKeepLeanMed & \ablKeepLeanMin--\ablKeepLeanMax \\
 & Lean memory & \ablKeepMemN & \ablKeepMemMed & \ablKeepMemMin--\ablKeepMemMax \\
 & clauses of $\Phi$ loaded & \ablKeepShareN & \ablKeepShareMed\% & \ablKeepShareMin--\ablKeepShareMax\% \\
Inlining & forward check time & \ablInlinePartBN & \ablInlinePartBMed & \ablInlinePartBMin--\ablInlinePartBMax \\
 & translator time & \ablInlineTransN & \ablInlineTransMed & \ablInlineTransMin--\ablInlineTransMax \\
All three shortening steps & forward check time & \ablAllPartBN & \ablAllPartBMed & \ablAllPartBMin--\ablAllPartBMax \\
\bottomrule
\end{tabular}
\end{table}

\paragraph{Ablation.}
In \cref{tab:ablation}, we measure the effect of each design choice (the instances and how we chose them are in \cref{app:provenance}).
Large effects come from leaving the deletions out of the certificate, which halves it in the median, and from the bit-parallel reverse check.
lrat-catcher can also check a refutation in the kernel alone, without native evaluation.
For the $4\times 4$ domino board (\kerFourAdds{} additions), the kernel takes \kerFourKernelS~s against \kerFourNativeMs~ms natively, and on the $6\times 6$ board it exceeds \kerLimitGB~GB.
For larger instances, we thus use native evaluation.

\section{Conclusion}\label{sec:conclusion}

We presented \tool, a tool that turns a CPOG certificate of an exact model counter into a Lean theorem about the formula as a term.
Combined with an encoding lemma, this gives a theorem about a count of objects.
The trusted base consists of Lean's kernel with the compiler and runtime that execute the checks.
\tool accepts any certificate in the fragment of the CPOG format (\cref{sec:parta}).
So any model counter that produces this format can serve as its front end; currently, this is d4 with cpog-gen.
The generator limits which counts we can prove (\cref{tab:funnel} in the appendix) and takes longer for the refutation than d4 for generating the count.

It would be interesting to extend a counter to write the refutation during its search.
This would avoid the need for a separate generator.
To scale the approach to even larger instances, one could split a count by cubes that the model counter chooses, with the sum proved as for the fibers in \cref{sec:burnside}.
To extend our approach to weighted counts, as used for the equivalence checking of quantum circuits~\cite{HuangCCCHLM26}, requires changing the validity check and the count theorem.
For projected counts, whose circuits SCPOG certifies, one would also need to change the reverse check.

\subsection*{Acknowledgments}
The authors used AI assistance for the implementation, for drafting and reviewing the text, and for creating the figures and tables; the authors take full responsibility for the content.
Stefan Szeider acknowledges the support of the Austrian Science Fund (FWF), grants \href{https://doi.org/10.55776/COE12}{10.55776/COE12} and \href{https://doi.org/10.55776/P36688}{10.55776/P36688}.

\bibliographystyle{splncs04}
\bibliography{references}

\begin{thebibliography}{10}
\providecommand{\url}[1]{\texttt{#1}}
\providecommand{\urlprefix}{URL }
\providecommand{\doi}[1]{https://doi.org/#1}

\bibitem{BermanK76}
Berman, J., K{\"{o}}hler, P.: Cardinalities of finite distributive lattices.
  Mitt. Math. Sem. Giessen  \textbf{121},  103--124 (1976),
  \url{https://oeis.org/A006356/a006356.pdf}

\bibitem{BeyersdorffFHHK26}
Beyersdorff, O., Fichte, J.K., Hecher, M., Hoffmann, T., Kasche, L.: The
  relative strength of {{\#}SAT} proof systems. J. Autom. Reason.
  \textbf{70}(2), ~12 (2026). \doi{10.1007/s10817-026-09757-w}

\bibitem{BeyersdorffHK25}
Beyersdorff, O., Hoffmann, T., Kasche, K.: Proof systems that tightly
  characterise model counting algorithms. Electronic Colloquium on
  Computational Complexity, Report TR25-127 (2025),
  \url{https://eccc.weizmann.ac.il/report/2025/127/}

\bibitem{BiereFFFFP24}
Biere, A., Faller, T., Fazekas, K., Fleury, M., Froleyks, N., Pollitt, F.:
  {CaDiCaL} 2.0. In: Gurfinkel, A., Ganesh, V. (eds.) Computer Aided
  Verification - 36th International Conference, {CAV} 2024, Montreal, QC,
  Canada, July 24-27, 2024, Proceedings, Part {I}. Lecture Notes in Computer
  Science, vol. 14681, pp. 133--152. Springer (2024).
  \doi{10.1007/978-3-031-65627-9\_7}

\bibitem{Boving25}
B{\"{o}}ving, H., Bhat, S., Cicolini, L., Keizer, A.C., Fr{\'{e}}not, L.,
  Mohamed, A., Stefanesco, L., Khan, H., Clune, J., Barrett, C.W., Grosser, T.:
  Interactive bitvector reasoning using verified bit-blasting. Proc. {ACM}
  Program. Lang.  \textbf{9}({OOPSLA2}),  3259--3285 (2025).
  \doi{10.1145/3763167}

\bibitem{BryantNAH23}
Bryant, R.E., Nawrocki, W., Avigad, J., Heule, M.J.H.: Certified knowledge
  compilation with application to verified model counting. In: Mahajan, M.,
  Slivovsky, F. (eds.) 26th International Conference on Theory and Applications
  of Satisfiability Testing, {SAT} 2023, Alghero, Italy, July 4-8, 2023.
  LIPIcs, vol.~271, pp. 6:1--6:20. Schloss Dagstuhl - Leibniz-Zentrum f{\"{u}}r
  Informatik (2023). \doi{10.4230/LIPICS.SAT.2023.6}

\bibitem{BryantNAH25}
Bryant, R.E., Nawrocki, W., Avigad, J., Heule, M.J.H.: Certified knowledge
  compilation with application to formally verified model counting. J. Artif.
  Intell. Res.  \textbf{82},  2057--2099 (2025). \doi{10.1613/JAIR.1.15958}

\bibitem{BryantTH25}
Bryant, R.E., Tan, Y.K., Heule, M.J.H.: Certifying projected knowledge
  compilation. In: Berg, J., Nordstr{\"{o}}m, J. (eds.) 28th International
  Conference on Theory and Applications of Satisfiability Testing, {SAT} 2025,
  Glasgow, Scotland, August 12-15, 2025. LIPIcs, vol.~341, pp. 8:1--8:22.
  Schloss Dagstuhl - Leibniz-Zentrum f{\"{u}}r Informatik (2025).
  \doi{10.4230/LIPICS.SAT.2025.8}

\bibitem{Capelli19}
Capelli, F.: Knowledge compilation languages as proof systems. In: Janota, M.,
  Lynce, I. (eds.) Theory and Applications of Satisfiability Testing - {SAT}
  2019 - 22nd International Conference, {SAT} 2019, Lisbon, Portugal, July
  9-12, 2019, Proceedings. Lecture Notes in Computer Science, vol. 11628, pp.
  90--99. Springer (2019). \doi{10.1007/978-3-030-24258-9\_6}

\bibitem{CapelliCMMB26}
Capelli, F., Choi, Y., Mengel, S., Mu{\~{n}}oz, M., den Broeck, G.V.: A
  canonical generalization of {OBDD}. In: Ignatiev, A., Szeider, S. (eds.) 29th
  International Conference on Theory and Applications of Satisfiability
  Testing, {SAT} 2026. LIPIcs, vol.~377, pp. 10:1--10:19. Schloss Dagstuhl -
  Leibniz-Zentrum f{\"{u}}r Informatik (2026). \doi{10.4230/LIPICS.SAT.2026.10}

\bibitem{CapelliLM21}
Capelli, F., Lagniez, J., Marquis, P.: Certifying top-down decision-{DNNF}
  compilers. In: Thirty-Fifth {AAAI} Conference on Artificial Intelligence,
  {AAAI} 2021, Thirty-Third Conference on Innovative Applications of Artificial
  Intelligence, {IAAI} 2021, The Eleventh Symposium on Educational Advances in
  Artificial Intelligence, {EAAI} 2021, Virtual Event, February 2-9, 2021. pp.
  6244--6253. {AAAI} Press (2021). \doi{10.1609/AAAI.V35I7.16776}

\bibitem{ChedeCS24}
Chede, S., Chew, L., Shukla, A.: Circuits, proofs and propositional model
  counting. In: Barman, S., Lasota, S. (eds.) 44th {IARCS} Annual Conference on
  Foundations of Software Technology and Theoretical Computer Science, {FSTTCS}
  2024, Gandhinagar, Gujarat, India, December 16-18, 2024. LIPIcs, vol.~323,
  pp. 18:1--18:23. Schloss Dagstuhl - Leibniz-Zentrum f{\"{u}}r Informatik
  (2024). \doi{10.4230/LIPICS.FSTTCS.2024.18}

\bibitem{Church65}
Church, R.: Enumeration by rank of the elements of the free distributive
  lattice with seven generators. Notices Amer. Math. Soc.  \textbf{12}(6), ~724
  (1965),
  \url{https://www.ams.org/journals/notices/196510/196510FullIssue.pdf},
  abstract 65T-447

\bibitem{Cruz-FilipeHHKS17}
Cruz{-}Filipe, L., Heule, M.J.H., Hunt, Jr., W.A., Kaufmann, M.,
  Schneider{-}Kamp, P.: Efficient certified {RAT} verification. In: de~Moura,
  L. (ed.) Automated Deduction - {CADE} 26 - 26th International Conference on
  Automated Deduction, Gothenburg, Sweden, August 6-11, 2017, Proceedings.
  Lecture Notes in Computer Science, vol. 10395, pp. 220--236. Springer (2017).
  \doi{10.1007/978-3-319-63046-5\_14}

\bibitem{DancoJCA25}
Danco, M., Janota, M., Codish, M., Ara{\'{u}}jo, J.J.: Complete symmetry
  breaking for finite models. In: Walsh, T., Shah, J., Kolter, Z. (eds.)
  Thirty-Ninth {AAAI} Conference on Artificial Intelligence, Thirty-Seventh
  Conference on Innovative Applications of Artificial Intelligence, Fifteenth
  Symposium on Educational Advances in Artificial Intelligence, {AAAI} 2025,
  Philadelphia, PA, USA, February 25 - March 4, 2025. pp. 11194--11202. {AAAI}
  Press (2025). \doi{10.1609/AAAI.V39I11.33217}

\bibitem{DarwicheM02}
Darwiche, A., Marquis, P.: A knowledge compilation map. J. Artif. Intell. Res.
  \textbf{17},  229--264 (2002). \doi{10.1613/JAIR.989}

\bibitem{FichteH23}
Fichte, J.K., Hecher, M.: {Model Counting Competition 2022}: Competition
  instances. Zenodo dataset (2023). \doi{10.5281/zenodo.10012860}

\bibitem{FichteH26}
Fichte, J.K., Hecher, M.: The model counting competitions 2021--2023. Artif.
  Intell.  \textbf{360},  104605 (2026). \doi{10.1016/J.ARTINT.2026.104605}

\bibitem{FichteHR22}
Fichte, J.K., Hecher, M., Roland, V.: Proofs for propositional model counting.
  In: Meel, K.S., Strichman, O. (eds.) 25th International Conference on Theory
  and Applications of Satisfiability Testing, {SAT} 2022, Haifa, Israel, August
  2-5, 2022. LIPIcs, vol.~236, pp. 30:1--30:24. Schloss Dagstuhl -
  Leibniz-Zentrum f{\"{u}}r Informatik (2022). \doi{10.4230/LIPICS.SAT.2022.30}

\bibitem{GallicchioCAH26}
Gallicchio, J., Codel, C.R., Avigad, J., Heule, M.J.H.: An end-to-end
  verification of {Keller}'s conjecture. In: Komendantskaya, E., Nipkow, T.
  (eds.) 17th International Conference on Interactive Theorem Proving, {ITP}
  2026, Lisbon, Portugal, July 26-29, 2026. LIPIcs, vol.~382, pp. 26:1--26:20.
  Schloss Dagstuhl - Leibniz-Zentrum f{\"{u}}r Informatik (2026).
  \doi{10.4230/LIPICS.ITP.2026.26}

\bibitem{GomesSS21}
Gomes, C.P., Sabharwal, A., Selman, B.: Model counting. In: Biere, A., Heule,
  M., van Maaren, H., Walsh, T. (eds.) Handbook of Satisfiability - Second
  Edition, Frontiers in Artificial Intelligence and Applications, vol.~336, pp.
  993--1014. {IOS} Press (2021). \doi{10.3233/FAIA201009}

\bibitem{HararyP73}
Harary, F., Palmer, E.M.: Graphical Enumeration. Academic Press, New York and
  London (1973)

\bibitem{HuangD07}
Huang, J., Darwiche, A.: The language of search. J. Artif. Intell. Res.
  \textbf{29},  191--219 (2007). \doi{10.1613/JAIR.2097}

\bibitem{HuangCCCHLM26}
Huang, W., Chareton, C., Chen, Y., Chung, K., Hsieh, M., Laarman, A., Mei, J.:
  Equivalence checking of quantum circuits via path-sum and weighted model
  counting. In: Junges, S., Katz, G. (eds.) Tools and Algorithms for the
  Construction and Analysis of Systems - 32nd International Conference, {TACAS}
  2026, Held as Part of the International Joint Conferences on Theory and
  Practice of Software, {ETAPS} 2026, Turin, Italy, April 11-16, 2026,
  Proceedings, Part {II}. Lecture Notes in Computer Science, vol. 16506, pp.
  419--439. Springer (2026). \doi{10.1007/978-3-032-22749-2\_21}

\bibitem{JohnsonT96}
Johnson, D.S., Trick, M.A. (eds.): Cliques, Coloring, and Satisfiability,
  Proceedings of a {DIMACS} Workshop, New Brunswick, New Jersey, {USA}, October
  11-13, 1993, {DIMACS} Series in Discrete Mathematics and Theoretical Computer
  Science, vol.~26. {DIMACS/AMS} (1996). \doi{10.1090/DIMACS/026}

\bibitem{Kery64}
K{\'e}ry, G.: {Ramsey} egy gr{\'a}felm{\'e}leti t{\'e}tel{\'e}r{\H{o}}l.
  Matematikai Lapok  \textbf{15},  204--224 (1964), in Hungarian

\bibitem{KieselE23}
Kiesel, R., Eiter, T.: Knowledge compilation and more with {SharpSAT}-{TD}. In:
  Marquis, P., Son, T.C., Kern{-}Isberner, G. (eds.) Proceedings of the 20th
  International Conference on Principles of Knowledge Representation and
  Reasoning, {KR} 2023, Rhodes, Greece, September 2-8, 2023. pp. 406--416
  (2023). \doi{10.24963/KR.2023/40}

\bibitem{KirchwegerMS26}
Kirchweger, M., Manrique, P., Szeider, S.: Formally verified graph generation
  with {SAT} modulo symmetries and {Lean}. In: Biere, A., Lutz, C., Negri, S.
  (eds.) Automated Reasoning - 13th International Joint Conference, {IJCAR}
  2026, Lisbon, Portugal, July 26-29, 2026, Proceedings, Part {I}. Lecture
  Notes in Computer Science, vol. 16688, pp. 117--135. Springer (2026).
  \doi{10.1007/978-3-032-32589-1\_8}

\bibitem{KorhonenJ21}
Korhonen, T., J{\"{a}}rvisalo, M.: Integrating tree decompositions into
  decision heuristics of propositional model counters (short paper). In:
  Michel, L.D. (ed.) 27th International Conference on Principles and Practice
  of Constraint Programming, {CP} 2021, Montpellier, France (Virtual
  Conference), October 25-29, 2021. LIPIcs, vol.~210, pp. 8:1--8:11. Schloss
  Dagstuhl - Leibniz-Zentrum f{\"{u}}r Informatik (2021).
  \doi{10.4230/LIPICS.CP.2021.8}

\bibitem{LagniezM17}
Lagniez, J., Marquis, P.: An improved decision-{DNNF} compiler. In: Sierra, C.
  (ed.) Proceedings of the Twenty-Sixth International Joint Conference on
  Artificial Intelligence, {IJCAI} 2017, Melbourne, Australia, August 19-25,
  2017. pp. 667--673. ijcai.org (2017). \doi{10.24963/IJCAI.2017/93}

\bibitem{LaiMY25}
Lai, Y., Meel, K.S., Yap, R.H.C.: Panini: An efficient and flexible knowledge
  compiler. In: Piskac, R., Rakamari{\'{c}}, Z. (eds.) Computer Aided
  Verification - 37th International Conference, {CAV} 2025, Zagreb, Croatia,
  July 23-25, 2025, Proceedings, Part {III}. Lecture Notes in Computer Science,
  vol. 15933, pp. 92--105. Springer (2025). \doi{10.1007/978-3-031-98682-6\_6}

\bibitem{LobbingW96}
L{\"{o}}bbing, M., Wegener, I.: The number of knight's tours equals
  33,439,123,484,294 --- counting with binary decision diagrams. Electron. J.
  Comb.  \textbf{3}(1), ~R5 (1996). \doi{10.37236/1229}

\bibitem{Lunnon71}
Lunnon, W.F.: The {IU} function: The size of a free distributive lattice. In:
  Welsh, D.J.A. (ed.) Combinatorial Mathematics and its Applications, pp.
  173--181. Academic Press, London (1971)

\bibitem{McCreeshNOT26}
McCreesh, C., Nordstr{\"{o}}m, J., Oertel, A., Tan, Y.K.: Proof logging for
  projected enumeration (and counting?) problems in {VeriPB}. In: Beldiceanu,
  N. (ed.) 32nd International Conference on Principles and Practice of
  Constraint Programming, {CP} 2026, Lisbon, Portugal, July 20-23, 2026.
  LIPIcs, vol.~379, pp. 43:1--43:21. Schloss Dagstuhl - Leibniz-Zentrum
  f{\"{u}}r Informatik (2026). \doi{10.4230/LIPICS.CP.2026.43}

\bibitem{McKay97}
McKay, B.D.: Knight's tours of an $8 \times 8$ chessboard. Tech. Rep.
  TR-CS-97-03, Department of Computer Science, Australian National University
  (1997), \url{https://oeis.org/A001230/a001230.pdf}

\bibitem{McKayW05}
McKay, B.D., Wanless, I.M.: On the number of {Latin} squares. Ann. Comb.
  \textbf{9}(3),  335--344 (2005). \doi{10.1007/s00026-005-0261-7}

\bibitem{MohamedMKBRQTB25}
Mohamed, A., Mascarenhas, T., Khan, H., Barbosa, H., Reynolds, A., Qian, Y.,
  Tinelli, C., Barrett, C.W.: lean-smt: An {SMT} tactic for discharging proof
  goals in {Lean}. In: Piskac, R., Rakamaric, Z. (eds.) Computer Aided
  Verification - 37th International Conference, {CAV} 2025, Zagreb, Croatia,
  July 23-25, 2025, Proceedings, Part {III}. Lecture Notes in Computer Science,
  vol. 15933, pp. 197--212. Springer (2025).
  \doi{10.1007/978-3-031-98682-6\_11}

\bibitem{MoreiraS26}
Moreira, W., Stubbs, J.: Sequencelib: {A} computational platform for
  formalizing the {OEIS} in {Lean}. CoRR  \textbf{abs/2601.11757} (2026).
  \doi{10.48550/ARXIV.2601.11757}

\bibitem{MouraU21}
de~Moura, L., Ullrich, S.: The {Lean} 4 theorem prover and programming
  language. In: Platzer, A., Sutcliffe, G. (eds.) Automated Deduction - {CADE}
  28 - 28th International Conference on Automated Deduction, Virtual Event,
  July 12-15, 2021, Proceedings. Lecture Notes in Computer Science, vol. 12699,
  pp. 625--635. Springer (2021). \doi{10.1007/978-3-030-79876-5\_37}

\bibitem{Nawrocki26}
Nawrocki, W.: Three certificates in computer-assisted mathematics: empty
  hexagons, model counting, and categorical semantics. Ph.D. thesis, Carnegie
  Mellon University, Department of Philosophy, Pittsburgh, PA, USA (2026),
  \url{https://voidma.in/assets/papers/2026nawrocki_three_certificates_computer_assisted_mathematics_empty_hexagons_model_counting_categorical_semantics.pdf}

\bibitem{Norton39}
Norton, H.W.: The $7 \times 7$ squares. Annals of Eugenics  \textbf{9}(3),
  269--307 (1939). \doi{10.1111/j.1469-1809.1939.tb02214.x}

\bibitem{Polya37}
P{\'{o}}lya, G.: Kombinatorische {Anzahlbestimmungen} f{\"{u}}r {Gruppen},
  {Graphen} und chemische {Verbindungen}. Acta Math.  \textbf{68},  145--254
  (1937). \doi{10.1007/BF02546665}

\bibitem{Radziszowski}
Radziszowski, S.P.: Small {Ramsey} numbers. Electron. J. Comb.  \textbf{Dynamic
  Survey DS1} (2026). \doi{10.37236/21}, revision 18 of April 24, 2026

\bibitem{Sade48}
Sade, A.: {\'E}num{\'e}ration des carr{\'e}s latins. {Application} au
  7\textsuperscript{\`eme} ordre. {Conjectures} pour les ordres sup{\'e}rieurs.
  Privately published, Marseille, 8 pp. (1948)

\bibitem{Sade51}
Sade, A.: An omission in {Norton}'s list of $7 \times 7$ squares. Ann. Math.
  Statist.  \textbf{22}(2),  306--307 (1951). \doi{10.1214/aoms/1177729654}

\bibitem{SharmaRSM19}
Sharma, S., Roy, S., Soos, M., Meel, K.S.: {GANAK:} {A} scalable probabilistic
  exact model counter. In: Kraus, S. (ed.) Proceedings of the Twenty-Eighth
  International Joint Conference on Artificial Intelligence, {IJCAI} 2019,
  Macao, China, August 10-16, 2019. pp. 1169--1176. ijcai.org (2019).
  \doi{10.24963/IJCAI.2019/163}

\bibitem{SubercaseauxNGC24}
Subercaseaux, B., Nawrocki, W., Gallicchio, J., Codel, C.R., Carneiro, M.,
  Heule, M.J.H.: Formal verification of the empty hexagon number. In: Bertot,
  Y., Kutsia, T., Norrish, M. (eds.) 15th International Conference on
  Interactive Theorem Proving, {ITP} 2024, Tbilisi, Georgia, September 9-14,
  2024. LIPIcs, vol.~309, pp. 35:1--35:19. Schloss Dagstuhl - Leibniz-Zentrum
  f{\"{u}}r Informatik (2024). \doi{10.4230/LIPICS.ITP.2024.35}

\bibitem{Szeider26a}
Szeider, S.: Streaming {LRAT} certificates into {Lean} theorems. CoRR
  \textbf{abs/2607.00815} (2026). \doi{10.48550/ARXIV.2607.00815}

\bibitem{TanHM21}
Tan, Y.K., Heule, M.J.H., Myreen, M.O.: cake{\_}lpr: Verified propagation
  redundancy checking in {CakeML}. In: Groote, J.F., Larsen, K.G. (eds.) Tools
  and Algorithms for the Construction and Analysis of Systems - 27th
  International Conference, {TACAS} 2021, Held as Part of the European Joint
  Conferences on Theory and Practice of Software, {ETAPS} 2021, Luxembourg
  City, Luxembourg, March 27 - April 1, 2021, Proceedings, Part {II}. Lecture
  Notes in Computer Science, vol. 12652, pp. 223--241. Springer (2021).
  \doi{10.1007/978-3-030-72013-1\_12}

\bibitem{TanYSMM24}
Tan, Y.K., Yang, J., Soos, M., Myreen, M.O., Meel, K.S.: Formally certified
  approximate model counting. In: Gurfinkel, A., Ganesh, V. (eds.) Computer
  Aided Verification - 36th International Conference, {CAV} 2024, Montreal, QC,
  Canada, July 24-27, 2024, Proceedings, Part {I}. Lecture Notes in Computer
  Science, vol. 14681, pp. 153--177. Springer (2024).
  \doi{10.1007/978-3-031-65627-9\_8}

\bibitem{Mathlib20}
{The mathlib Community}: The {Lean} mathematical library. In: Blanchette, J.,
  Hritcu, C. (eds.) Proceedings of the 9th {ACM} {SIGPLAN} International
  Conference on Certified Programs and Proofs, {CPP} 2020, New Orleans, LA,
  USA, January 20-21, 2020. pp. 367--381. {ACM} (2020).
  \doi{10.1145/3372885.3373824}

\bibitem{OEIS-A000170}
{The OEIS Foundation Inc.}: Sequence {A000170}: Number of ways of placing $n$
  nonattacking queens on an $n \times n$ board. The On-Line Encyclopedia of
  Integer Sequences (2026), \url{https://oeis.org/A000170}

\bibitem{OEIS-A004003}
{The OEIS Foundation Inc.}: Sequence {A004003}: Number of domino tilings (or
  dimer coverings) of a $2n \times 2n$ square. The On-Line Encyclopedia of
  Integer Sequences (2026), \url{https://oeis.org/A004003}

\bibitem{OEIS-A006506}
{The OEIS Foundation Inc.}: Sequence {A006506}: Number of $n \times n$ binary
  matrices with no 2 adjacent 1's; also number of independent vertex sets in an
  $n \times n$ grid. The On-Line Encyclopedia of Integer Sequences (2026),
  \url{https://oeis.org/A006506}

\bibitem{OEIS-A006785}
{The OEIS Foundation Inc.}: Sequence {A006785}: Number of triangle-free graphs
  on $n$ vertices. The On-Line Encyclopedia of Integer Sequences (2026),
  \url{https://oeis.org/A006785}

\bibitem{OEIS-A303210}
{The OEIS Foundation Inc.}: Sequence {A303210}: Number of total dominating sets
  in the $n \times n$ torus grid graph. The On-Line Encyclopedia of Integer
  Sequences (2026), \url{https://oeis.org/A303210}

\bibitem{OEIS-A358391}
{The OEIS Foundation Inc.}: Sequence {A358391}: The number of antichains in the
  {Kreweras} lattice of non-crossing set partitions of an $n$-element set. The
  On-Line Encyclopedia of Integer Sequences (2026),
  \url{https://oeis.org/A358391}

\bibitem{OEIS-A396728}
{The OEIS Foundation Inc.}: Sequence {A396728}: Number of antichains of subsets
  of an $n$-element set ({Sperner} families) that do not contain both a set and
  its complement. The On-Line Encyclopedia of Integer Sequences (2026),
  \url{https://oeis.org/A396728}

\bibitem{OEIS}
{The OEIS Foundation Inc.}: {The On-Line Encyclopedia of Integer Sequences}
  (2026), \url{https://oeis.org}

\bibitem{Valiant79}
Valiant, L.G.: The complexity of enumeration and reliability problems. {SIAM}
  J. Comput.  \textbf{8}(3),  410--421 (1979). \doi{10.1137/0208032}

\bibitem{Wells67}
Wells, M.B.: The number of {Latin} squares of order eight. J. Comb. Theory
  \textbf{3}(1),  98--99 (1967). \doi{10.1016/S0021-9800(67)80021-8}

\bibitem{WetzlerHH14}
Wetzler, N., Heule, M., Hunt, Jr., W.A.: {DRAT}-trim: Efficient checking and
  trimming using expressive clausal proofs. In: Sinz, C., Egly, U. (eds.)
  Theory and Applications of Satisfiability Testing - {SAT} 2014 - 17th
  International Conference, Held as Part of the Vienna Summer of Logic, {VSL}
  2014, Vienna, Austria, July 14-17, 2014. Proceedings. Lecture Notes in
  Computer Science, vol.~8561, pp. 422--429. Springer (2014).
  \doi{10.1007/978-3-319-09284-3\_31}

\end{thebibliography}

\appendix
\raggedbottom
\makeatletter\setlength{\@fptop}{0pt}\makeatother

\section{Additional Tables and Full Statements}\label[appendix]{sec:appendix}

\subsection{Resources of the Theorems about Mathematical Objects}\label[appendix]{app:showcases}

In \cref{tab:app-showcases} we give the resources of the theorems of \cref{tab:showcases}.
For the Latin squares we ran one Lean process per certificate and one for the assembled theorem, and we report the largest peak memory of these processes.
The largest certificate of order~8 has \scLatinEightCertMaxGB~GB.

\begin{table}[!htbp]
\centering
\caption{Resources of the theorems of \cref{tab:showcases}, in the same groups. ``Summands'' is the number of certified counts, as in \cref{tab:showcases}. ``Size'' is the total size of the certificates in GB as written by the patched cpog-gen without deletions; for a count of zero it is the size of the LRAT refutation. ``d4'' is the time of d4 and ``Gen.'' the time of cpog-gen (for a count of zero, of the SAT solver that refutes the formula), in seconds and summed over the certificates. ``Lean'' is the Lean time in seconds as in \cref{tab:showcases}. ``Mem.'' is the peak memory of the Lean process in GB, the largest one if there are several. ``Axioms'' is the number of native-evaluation axioms that \texttt{\#print axioms} lists for the theorem: one per embedded and two per streamed certificate.}
\label{tab:app-showcases}
\footnotesize
\setlength{\tabcolsep}{3pt}
\begin{tabular}{@{}>{\raggedright\arraybackslash}p{3.5cm}rrrrrrr@{}}
\toprule
Theorem & Summands & Size & d4 & Gen. & Lean & Mem. & Axioms\\
\midrule
\multicolumn{8}{@{}l}{\emph{Composition of certified counts}} \\
\hangindent=1em\hangafter=0 Triangle-free graphs & 42 & 2.3 & 25.7 & 80.8 & \scTriTenLean & 9.2 & 44 \\
\hangindent=1em\hangafter=0 Latin squares, order 7 & 4 & 0.70 & 3.8 & 26.6 & \scLatinSevenLean & 3.9 & 8 \\
\hangindent=1em\hangafter=0 Latin squares, order 8 & 737 & 7{,}929 & 36{,}376 & 486{,}490 & \scLatinEightLean & 21.6 & \scLatinEightNativeAxioms \\
\addlinespace
\multicolumn{8}{@{}l}{\emph{One count as an end-to-end theorem}} \\
\hangindent=1em\hangafter=0 Dedekind's $M(7)$ & 1 & 1.7 & 26.2 & 116.9 & \scDedekindSevenLean & 9.3 & 2 \\
\hangindent=1em\hangafter=0 A358391(6) & 1 & 2.9 & 62.7 & 206.7 & \scKrewerasSixLean & 13.4 & 2 \\
\hangindent=1em\hangafter=0 A396728(7) & 1 & 7.2 & 63.7 & 271.8 & \scSpernerSevenLean & 18.2 & 2 \\
\hangindent=1em\hangafter=0 A303210(9) & 1 & 4.3 & 47.6 & 167.2 & \scTotDomNineLean & 11.8 & 2 \\
\hangindent=1em\hangafter=0 A303210(10) & 1 & 0.24 & 3.3 & 9.0 & \scTotDomTenLean & 3.9 & 2 \\
\addlinespace
\multicolumn{8}{@{}l}{\emph{A count as a hypothesis}} \\
\hangindent=1em\hangafter=0 $(3,5)$-Ramsey graph & 113 & 3.4 & 30.3 & 660.5 & \scRamseyThirteenLean & 5.4 & 115 \\
\bottomrule
\end{tabular}
\end{table}

\subsection{Validation Families}\label[appendix]{app:families}

In \cref{tab:app-families} we list the instances of the validation families of \cref{sec:eval} with their counts and the resources of the pipeline, and we prove each count as a theorem about the objects with a streamed refutation.
The counts agree with the sequences A004003 (domino tilings), A000170 (queens) and A006506 (independent sets of grids) of the OEIS~\cite{OEIS-A004003,OEIS-A000170,OEIS-A006506}; for the graphs, the counters ganak~\cite{SharmaRSM19} and SharpSAT-TD~\cite{KorhonenJ21} give the same counts.
The graphs games120 and homer are graph coloring benchmarks of the DIMACS challenge~\cite{JohnsonT96}, and we generated the cubic graphs at random with a fixed seed.

\begin{table}[!htbp]
\centering
\caption{The validation families by instance: the number of objects, the size of the certificate in MB as written by the patched cpog-gen without deletions, the generation (``Gen.'') and the Lean time in seconds, and the peak memory (``Mem.'') of the Lean process in GB.}
\label{tab:app-families}
\footnotesize
\setlength{\tabcolsep}{4pt}
\begin{tabular}{@{}lrrrrr@{}}
\toprule
Instance & Count & Size & Gen. & Lean & Mem.\\
\midrule
\multicolumn{6}{@{}l}{\emph{Domino tilings}} \\
\quad $12\times 12$ board & \ensuremath{5.31\cdot 10^{16}} & 500.1 & 225.4 & 20.1 & 3.6 \\
\addlinespace
\multicolumn{6}{@{}l}{\emph{Queens}} \\
\quad $n=12$ & 14{,}200 & 406.3 & 73.9 & 12.9 & 3.5 \\
\quad $n=13$ & 73{,}712 & 2{,}846 & 461.2 & 56.0 & 4.5 \\
\quad $n=14$ & 365{,}596 & 18{,}194 & 2{,}741 & 302.5 & 9.3 \\
\addlinespace
\multicolumn{6}{@{}l}{\emph{Independent sets of grids}} \\
\quad $14\times 14$ grid & \ensuremath{3.39\cdot 10^{35}} & 327.9 & 36.4 & 23.1 & 5.0 \\
\quad $16\times 16$ grid & \ensuremath{1.84\cdot 10^{46}} & 3{,}549 & 421.8 & 136.7 & 20.8 \\
\addlinespace
\multicolumn{6}{@{}l}{\emph{Independent sets of graphs}} \\
\quad hypercube $Q_{6}$, 64 vertices & \ensuremath{1.98\cdot 10^{10}} & 68.1 & 3.7 & 11.9 & 3.5 \\
\quad random cubic, 100 vertices & \ensuremath{8.28\cdot 10^{18}} & 117.8 & 11.2 & 12.8 & 4.0 \\
\quad random cubic, 120 vertices & \ensuremath{5.00\cdot 10^{22}} & 1{,}076 & 101.6 & 39.7 & 8.3 \\
\quad games120, 120 vertices & \ensuremath{3.05\cdot 10^{13}} & 860.6 & 88.3 & 43.1 & 7.5 \\
\quad homer, 561 vertices & \ensuremath{4.99\cdot 10^{123}} & 32.7 & 6.2 & 12.4 & 3.4 \\
\bottomrule
\end{tabular}
\end{table}

\subsection{The Competition Set}\label[appendix]{app:competition}

Bryant et al.~\cite{BryantNAH25} report that d4 completes 123 of the 180 formulas within \funLimitDFour~s.
They report further that their toolchain fully verifies 111 of the 123, using the unverified checker cpog-check and a limit of 10,000~s for generation, check and count together.
They run their other experiments on the 90 formulas on which d4 takes under 1,000~s and whose POG has at most $10^7$ defining clauses.
In \cref{tab:funnel} we count the formulas by stage and outcome, and in \cref{tab:app-failed} we name those that did not become theorems (track1 is the unweighted track, track2 the weighted one).

\begin{table}[!htbp]
\centering
\caption{The formulas of the competition set by pipeline stage: the formulas that reached each stage and their outcomes, with the time limit of the stage and \funLimitMem~GB of memory throughout. Only the checker can reject, and ``Passed'' on its row means proved.}
\label{tab:funnel}
\footnotesize
\setlength{\tabcolsep}{3pt}
\begin{tabular}{@{}>{\raggedright\arraybackslash}p{3.1cm}rrrrrr@{}}
\toprule
Stage & Reached & Rejected & Timeout & Out of memory & Crash & Passed\\
\midrule
d4 (\funLimitDFour~s) & \funDFourEntered & -- & \funDFourTimeout & \funDFourOom & \funDFourCrash & \funDFourPassed \\
Generation (\funLimitGen~s) & \funGenEntered & -- & \funGenTimeout & \funGenOom & \funGenCrash & \funGenPassed \\
Translation (\funLimitGen~s) & \funTransEntered & -- & \funTransTimeout & \funTransOom & \funTransCrash & \funTransPassed \\
Lean, theorem stage (\funLimitLean~s) & \funLeanEntered & \funLeanRejected & \funLeanTimeout & \funLeanOom & \funLeanCrash & \funLeanPassed \\
\bottomrule
\end{tabular}
\end{table}
cpog-gen failed on 26 formulas.
Of the 25 for which it wrote the POG before it failed, 20 have a POG with more than $10^7$ defining clauses.
On the four that exceeded the memory limit, d4 took more than 2,800~s.
In a rerun on 2026-10-11 with a generation limit of 10,000~s, \rgGenOk{} of the \rgFailN{} formulas generate a certificate, and \rgThm{} become theorems within the Lean limit of \funLimitLean~s (all \rgThmLeanTenK{} with 10,000~s).
We also ran each formula with one SAT call per subgraph below tree size~$10^6$, as Bryant et al. describe their generator.
This mode succeeded only where the structural mode did.
The four formulas that exceeded 64~GB fail also with 128~GB.

\begin{table}[!htbp]
\centering
\caption{The formulas of the competition set that did not become theorems, by the stage and the outcome of \cref{tab:funnel}; $n$ is the number of formulas.}
\label{tab:app-failed}
\footnotesize
\setlength{\tabcolsep}{4pt}
\begin{tabular}{@{}llr>{\raggedright\arraybackslash}p{7cm}@{}}
\toprule
Stage & Outcome & $n$ & Formulas\\
\midrule
d4 & timeout & 57 & track1\_001, track1\_003, track1\_061, track1\_067, track1\_085, track1\_105, track1\_111, track1\_115, track1\_131, track1\_137, track1\_139, track1\_159, track1\_161, track1\_163, track1\_165, track1\_169, track1\_173, track1\_175, track1\_177, track1\_179, track1\_181, track1\_183, track1\_185, track1\_187, track1\_189, track1\_191, track1\_193, track1\_195, track1\_197, track1\_199, track2\_003, track2\_065, track2\_069, track2\_071, track2\_075, track2\_083, track2\_099, track2\_105, track2\_113, track2\_135, track2\_141, track2\_145, track2\_151, track2\_153, track2\_157, track2\_163, track2\_171, track2\_173, track2\_175, track2\_177, track2\_181, track2\_185, track2\_187, track2\_189, track2\_191, track2\_193, track2\_195 \\
Generation & timeout & 22 & track1\_053, track1\_071, track1\_075, track1\_089, track1\_099, track1\_107, track1\_119, track1\_123, track1\_125, track1\_127, track1\_129, track1\_143, track1\_147, track1\_155, track2\_001, track2\_055, track2\_059, track2\_119, track2\_129, track2\_161, track2\_179, track2\_199 \\
Generation & out of memory & 4 & track1\_151, track2\_123, track2\_143, track2\_169 \\
Lean & timeout & 1 & track1\_083 \\
\bottomrule
\end{tabular}
\end{table}

\subsection{The Smallest Miscount of cpog-check}\label[appendix]{app:miscount}

\begin{sloppypar}
Of the \cmpN{} certificates of the comparison with CPOG's checkers, cpog-check reported a wrong count on \cmpCcWrong{} (\cref{fig:compare}).
The one with the smallest count is the instance ``\miscountFormula'' of \cref{tab:app-families}.
On it, cpog-check reported \miscountPrinted{}, while we prove \miscountOurs{}, the count that d4 and cpog-lean-check also give.
cpog-check computes every count with rational numbers whose denominators are products of powers of 2 and 5.
The fault is in the normalization of these numbers (the source file \texttt{q25.c}).
The normalization divides a number by the largest power of 2 or 5 that divides its lowest digit in base $10^9$, which need not divide the number, and discards the remainder.
We changed one line so that the power is at most the ninth.
With this change, cpog-check reported the correct count on every certificate we retested, among them the \cmpCcWrong{}.
We reported the fault to the developers of CPOG with the one-line fix (issue~3 of github.com/rebryant/cpog).
\end{sloppypar}

\subsection{Constructed Certificates and the Differential Test}\label[appendix]{app:adversarial}

For the tests of \cref{sec:trust} we count in \cref{tab:app-adversarial} the certificates that the checker rejected, by the first condition that fails.
The condition of the fragment of \cref{sec:parta} that neither an argument nor the root is a negated node is tested by the reverse check and not by the validity check, since the implication masks of the reverse check are defined for such POGs only.
For the differential test, d4 and the patched cpog-gen generated certificates from small formulas, and we mutated each in one place.
We checked them with the refutation in text format (embedded and streamed) and with the defaults of \cref{sec:design} (trimming, keep mask, inlining and the binary format), in a run of 2026-10-07 on the code of the artifact (\cref{tab:app-provenance}).
Every certificate that the checker accepted has the count we obtain by enumeration.

\begin{table}[!htbp]
\centering
\caption{Certificates rejected by the checker, counted at the first condition that fails. The rows are in the order in which the checker tests the conditions, with the reverse check before the forward check. The rows ``universe'' are the two conditions on the universe that the checker tests before the three checks (\cref{sec:contract}). ``Constructed'' are the certificates we built by hand, one or more per condition that violate it, and valid certificates of edge cases (an empty formula, no variables, tautologies only). ``Generated'' are the certificates of the differential test on small formulas, unmutated or with one mutation each; we do not count the mutants that the translator rejected. In the last two rows we count the accepted certificates and all certificates.}
\label{tab:app-adversarial}
\footnotesize
\setlength{\tabcolsep}{4pt}
\begin{tabular}{@{}l>{\raggedright\arraybackslash}p{6.4cm}rr@{}}
\toprule
Check & Condition & Constructed & Generated\\
\midrule
universe & every variable of $F$ lies in the universe & 1 & 0 \\
 & the node table is over the $V$ variables of the theorem & 1 & 0 \\
\addlinespace
validity & every argument is an earlier variable & 2 & 2 \\
 & the arguments of a product have disjoint supports & 3 & 739{,}166 \\
 & every exclusion hint cites a defining clause of an earlier node & 1 & 31{,}246 \\
 & the exclusion of a sum follows by unit propagation from its hints & 3 & 1{,}475{,}770 \\
 & the root is declared & 1 & 0 \\
\addlinespace
reverse & neither an argument nor the root is a negated node & 4 & 542{,}892 \\
 & the root implies every clause of $F$ that is not a tautology & 4 & 2{,}493{,}640 \\
\addlinespace
forward & the refutation is well-formed LRAT & 1 & 296{,}388 \\
 & every addition follows from its hints by unit propagation, and the last is the empty clause & 3 & 5{,}010{,}365 \\
\midrule
\multicolumn{2}{@{}l}{accepted, each with the count that enumeration gives} & 22 & 1{,}553{,}855 \\
\multicolumn{2}{@{}l}{total} & 46 & 12{,}143{,}324 \\
\bottomrule
\end{tabular}
\end{table}

\subsection{The Bit-Parallel Reverse Check}\label[appendix]{app:partc}

\Cref{alg:partc} is the reverse check of \cref{sec:partc} for one batch of clauses, as the checker executes it.

\begin{algorithm}[!htbp]
\caption{The reverse check for one batch of clauses $D_0,\dots,D_{k-1}$, $k\le 63$. Implication masks are nonnegative integers read as bit sets. For an argument~$a$, $\mathit{mask}(a)$ is $\mathit{occ}[a]$ if $a$ is a literal of an input variable and $m[a]$ if $a$ is a node.}
\label{alg:partc}
\footnotesize
\begin{algorithmic}[1]
\Require nodes $u_1,\dots,u_n$ in the order of their declaration, root~$r$, array $\mathit{occ}$ with all entries~$0$
\ForAll{$j<k$ and literals~$a$ of~$D_j$} $\mathit{occ}[a]\gets \mathit{occ}[a]\mathbin{|}2^j$ \EndFor
\For{$i=1,\dots,n$}
  \If{$u_i$ is a product with the set~$A$ of arguments} $m[u_i]\gets$ bitwise or of $\mathit{mask}(a)$ for $a\in A$ \EndIf
  \If{$u_i$ is a sum of $a$ and~$b$} $m[u_i]\gets \mathit{mask}(a)\mathbin{\&}\mathit{mask}(b)$ \EndIf
  \Statex \hspace{\algorithmicindent}$\triangleright$ invariant: if bit~$j$ of $\mathit{mask}(a)$ is set, then $a$ implies~$D_j$, for every literal and every node~$a$ seen so far
\EndFor
\State $\mathit{ok}\gets$ bit~$j$ of $\mathit{mask}(r)$ is set for every $j<k$ with $D_j$ not tautological
\ForAll{$j<k$ and literals~$a$ of~$D_j$} $\mathit{occ}[a]\gets 0$ \EndFor
\State \Return $\mathit{ok}$
\end{algorithmic}
\end{algorithm}

\subsection{Soundness Theorems}\label[appendix]{app:soundness}

In \cref{lst:soundness} we state the soundness theorems of \cref{sec:design} as they appear in the sources, and \texttt{\#print axioms} lists only propositional extensionality, choice and quotient soundness for each of them.
The checker and the LRAT checker are precompiled libraries, loaded as plugins; we build the definitions the kernel checks and their C code from the same source in one build, and Lean does not check at load time that the two agree.

\begin{lstlisting}[caption={The soundness theorems of \cref{sec:design} as stated in the sources (proofs omitted): the validity check, the forward implication from a refutation, the reverse check, the composition over one POG, the soundness of the whole check for an embedded and for a streamed refutation, the lemma of the keep mask, the soundness of lrat-catcher's streamed checker, which quantifies over every feed (the function that delivers the refutation in blocks), and the generic family lemma for a streamed refutation, which the fiber lemma behind \cref{lst:latin} instantiates. \lean{WF} is validity, \lean{CnfWF} the condition that every variable of $F$ lies in the universe, \lean{countRoot} the number $c(r)\cdot 2^{\,V-|S(r)|}$, and \lean{fwdCnf} and \lean{fwdCnfOf} are $\Phi$, from the parsed POG and from the certificate text.},label=lst:soundness]
theorem checkPog_sound {c : PogCert} {N : Nat} (h : checkPog c = some N) :
    c.pog.WF ∧ c.root.1 < c.pog.numVars ∧ N = c.pog.countRoot c.root
theorem fwd_of_unsat {F : CNF Nat} {P : Pog} {r : Literal Nat}
    (h : (fwdCnf F P r).Unsat) : Fwd F P r
theorem checkRevKillFast_sound {P : Pog} (hP : P.WF) {F : CNF Nat} {r : Literal Nat}
    (hr : r.1 < P.numVars) (h : checkRevKillFast P.V P F r = true) : RevEval F P r
theorem card_models_eq_countRoot {V : ℕ} {F : CNF ℕ} {P : Pog} (hF : CnfWF V F)
    (hP : P.WF) (hV : P.V = V) {r : Literal ℕ} (hr : r.1 < P.numVars)
    (fwd : Fwd F P r) (rev : RevEval F P r) :
    Fintype.card {a : Fin V → Bool // F.eval (ext a) = true} = P.countRoot r
theorem checkAll_sound {V : Nat} {F : CNF Nat} {pogStr lratStr : String} {N : Nat}
    (h : checkAll V F pogStr lratStr N = true) :
    Fintype.card {a : Fin V → Bool // F.eval (ext a) = true} = N
theorem checkAllAC_sound {V : Nat} {F : CNF Nat} {pogStr : String} {N : Nat}
    (h : checkAllAC V F pogStr N = true) (hB : (fwdCnfOf F pogStr).Unsat) :
    Fintype.card {a : Fin V → Bool // F.eval (ext a) = true} = N
theorem fwdCnfOf_unsat_of_sel {F : CNF Nat} {pogStr keepStr : String}
    (h : (fwdCnfOfSel F pogStr keepStr).Unsat) : (fwdCnfOf F pogStr).Unsat
theorem LRATCatcher.checkStreamCnf_sound (cnf : CNF Nat)
    (feed : Nat → Option (Array IntAction)) (fuel : Nat)
    (h : checkStreamCnf cnf feed fuel = true) : cnf.Unsat
theorem Family.card_of_checkAC {X : Type*} {V : ℕ} {F : CNF ℕ}
    (e : {a : Fin V → Bool // F.eval (ext a) = true} ≃ X) {pogStr : String} {N : ℕ}
    (h : checkAllAC V F pogStr N = true) (hB : (fwdCnfOf F pogStr).Unsat) : Nat.card X = N
\end{lstlisting}
 
\subsection{Provenance of the Measurements}\label[appendix]{app:provenance}

One script writes every number of the paper from the result files of the runs in \cref{tab:app-provenance}.
The artifact contains these files.
In the run of 2026-10-02 we used the tools of \cref{tab:app-tools}.
For the rows measured in other runs we name the differences in \cref{tab:app-provenance}.
cpog-gen derives the refutation node by node, by unit propagation where that suffices and otherwise by a call of CaDiCaL.

\begin{table}[!htbp]
\centering
\caption{The tools of the run of 2026-10-02. From CPOG we used cpog-gen with our two patches, the unpatched cpog-gen for the full certificates, and cpog-check and cpog-lean-check.}
\label{tab:app-tools}
\footnotesize
\setlength{\tabcolsep}{4pt}
\begin{tabular}{@{}lll@{}}
\toprule
Tool & Version & Source\\
\midrule
Lean & 4.30.0 & \\
Mathlib & commit c5ea003 & \\
lrat-catcher & 0.2.0 & \\
d4 & commit 333370c & github.com/crillab/d4\\
CPOG & release 1.2.0, commit a97ed85 & github.com/rebryant/cpog\\
CaDiCaL~\cite{BiereFFFFP24} & 3.0.1 & \\
drat-trim~\cite{WetzlerHH14} & commit 2e3b2dc & github.com/marijnheule/drat-trim\\
\bottomrule
\end{tabular}
\end{table}

We call the patched cpog-gen as \texttt{cpog-gen-fwd -F -T 500 -G}.
With \texttt{-F}, from our first patch, it omits the deletions, and with \texttt{-G} and \texttt{-T 500}, from our second patch, it frees proof clauses that no later step reads and shortens with drat-trim every solver proof that has at least 500 clauses.

Memory limits are enforced by the job scheduler of the cluster, and peak memory is the largest amount of main memory the process occupied, as reported by the operating system; both are in GB of $2^{30}$ bytes.
Both checkers of a comparison run one after the other on the same machine.

In the comparison we omit the 14 proved competition formulas with the longest generation times.
In a run before 2026-10-02, three of them had exceeded the 64~GB memory limit of the cpog-lean-check run; one of these three has the largest certificate of the set.
On the largest certificate of the comparison, \cmpBigInstance{} with \cmpBigFullGB~GB, cpog-lean-check takes \cmpBigVc~s and \cmpBigVcMem~GB, and the Lean process takes \cmpBigOurLean~s and \cmpBigOurMem~GB.

In \cref{tab:ablation} we exclude \ablPartcPcCensored{} sweeps that exceeded 1,800~s.
For the generation time we use the instances on which both variants of cpog-gen take at least a second.
We measured trimming and the binary format together.
For the bit-parallel reverse check of \cref{tab:ablation} we used the eleven instances of \cref{tab:app-families} and the six competition formulas track1\_063, track1\_149, track1\_171, track2\_081, track2\_089 and track2\_095, on which an earlier version of our reverse check had been slow.
For trimming with the binary format, the keep mask and inlining we took 13- and 14-queens, the $16\times 16$ grid, A396728(7), and track2\_011, track2\_087 and track2\_095, whose certificates we had stored from runs before 2026-10-02.
We also compared streaming the refutation with embedding it in the theorem file, on the eleven instances whose refutation in text format has at most 132~MB (embedded with \texttt{+embed} above the default limit of 64~MB), namely the $8\times 8$ and $10\times 10$ boards, 10- and 11-queens, the $10\times 10$ and $12\times 12$ grids, $Q_6$, random cubic~100, homer, track1\_011 and track2\_007.
\Cref{tab:app-stream} has the result in the format of \cref{tab:ablation}.

\begin{table}[!htbp]
\centering
\caption{Embedding the refutation in the theorem file against streaming it, in the format of \cref{tab:ablation}: the median and the range over the instances of the ratio of the check time with the refutation embedded to the check time with it streamed, once for the same refutation in text format and once with the keep mask, inlining and the binary format, which are available only for a streamed refutation.}
\label{tab:app-stream}
\footnotesize
\setlength{\tabcolsep}{4pt}
\begin{tabular}{@{}llrrr@{}}
\toprule
Design choice & Metric & Instances & Median & Range\\
\midrule
Streamed refutation & check time, same proof & \ablEmbedSameN & \ablEmbedSameMed & \ablEmbedSameMin--\ablEmbedSameMax \\
 & check time, with mask & \ablEmbedDefN & \ablEmbedDefMed & \ablEmbedDefMin--\ablEmbedDefMax \\
\bottomrule
\end{tabular}
\end{table}

\begin{table}[!htbp]
\centering
\caption{Provenance of the measured numbers: where each result is reported, the data behind it, the date of the run and the sources where they differ from the run of 2026-10-02, the machine class (cpu01: AMD EPYC 9554P; b4: Intel Xeon E5-2640~v4; b5: AMD EPYC 7402; gpu01: AMD EPYC 9274F) and the limits. We took every measurement once, except for the comparison of streaming with embedding and for the bit-parallel reverse check below 60~s, which we measured three times.}
\label{tab:app-provenance}
\footnotesize
\setlength{\tabcolsep}{3pt}
\begin{tabular}{@{}>{\raggedright}p{1.9cm}>{\raggedright}p{2.9cm}>{\raggedright}p{2.6cm}>{\raggedright}p{1.3cm}>{\raggedright\arraybackslash}p{2.5cm}@{}}
\toprule
Reported in & Data & Run & Machine & Limits\\
\midrule
\cref{tab:showcases}, \cref{tab:app-showcases} & the runs of the theorems (the five single counts without inlining); order~8: 737 runs, one per certificate, and one for the assembled theorem & 2026-10-02; order~8 to 2026-10-04, same pipeline & cpu01; order~8 b5 & 64~GB, 4~h; order~8 20 or 48~GB per certificate\\
\cref{tab:funnel}, \cref{tab:app-failed} & 180 runs of the whole pipeline & 2026-10-02 & b5 & d4 and generation \funLimitGen~s, Lean \funLimitLean~s, \funLimitMem~GB\\
\cref{fig:compare}, \cref{tab:bands}, \cref{app:miscount}, \cref{tab:ablation} deletions & 91 runs with both variants of cpog-gen and both CPOG checkers in one job & 2026-10-02 & b4 (89), b5 (2) & each checker \funLimitLean~s; 32, 48 or 64~GB\\
\cref{fig:compare}, unshortened refutation & 91 runs as above, plus a Lean run on the unshortened refutation & 2026-10-05 & b4 & as above\\
\cref{tab:ablation} bit-parallel reverse check & 17 instances, the reverse check alone & 2026-10-01; the top-down evaluation of that day & cpu01 & 1,800~s per check\\
\cref{tab:ablation} trimmed, keep mask, inlining & 7 instances, four translator settings & 2026-10-02; certificates of runs before 2026-10-02 & cpu01 & 32 or 64~GB, 2~h\\
\cref{tab:app-stream} & 11 instances, embedded and streamed & 2026-10-04 & cpu01 & 48~GB, 1,800~s\\
\cref{sec:eval}, rerun of the generation failures & \rgFailN{} formulas, two generator modes & 2026-10-11 & b5, cpu01 & generation 10,000~s; 64 or 128~GB\\
\cref{sec:eval}, kernel alone & domino boards $4\times 4$ to $6\times 6$ & 2026-10-04 & cpu01 & 64~GB, 3,600~s\\
\cref{tab:app-families} & 11 runs of the whole pipeline & 2026-10-02 & cpu01 & none\\
\cref{tab:app-adversarial} & constructed certificates; differential test of 12.1 million certificates & 2026-10-07, commit d097449; constructed certificates of 2026-10-11 with the shipped tools & gpu01; b4 machine for the constructed certificates & 20~s per run of each tool\\
\bottomrule
\end{tabular}
\end{table}
 
\end{document}